\documentclass[12pt,letterpaper]{article}

\global\arraycolsep=1pt
\usepackage{amssymb}
\usepackage{color}
\usepackage{graphicx}
\usepackage{tipa}
\usepackage{lscape}
\usepackage{amsmath}

\font\cyr=wncyr10 
\newcommand{\Lov}{\operatorname{\mbox{\cyr L}}}

\numberwithin{equation}{section}

\graphicspath{{Fig_reduced/}}

\newcommand{\bC}{\ensuremath{\mathbb{C}}}

\newcommand{\bH}{\ensuremath{\mathbb{H}}}

\newcommand{\bP}{\ensuremath{\mathbb{P}}}

\newcommand{\bR}{\ensuremath{\mathbb{R}}}

\newcommand{\bZ}{\ensuremath{\mathbb{Z}}}

\newcommand{\scA}{\ensuremath{\mathcal{A}}}

\newcommand{\scE}{\ensuremath{\mathcal{E}}}
\newcommand{\scF}{\ensuremath{\mathcal{F}}}
\newcommand{\scG}{\ensuremath{\mathcal{G}}}
\newcommand{\scH}{\ensuremath{\mathcal{H}}}

\newcommand{\scL}{\ensuremath{\mathcal{L}}}

\newcommand{\scN}{\ensuremath{\mathcal{N}}}
\newcommand{\scO}{\ensuremath{\mathcal{O}}}

\newcommand{\scQ}{\ensuremath{\mathcal{Q}}}

\newcommand{\scS}{\ensuremath{\mathcal{S}}}

\newcommand{\scW}{\ensuremath{\mathcal{W}}}

\newcommand{\bea}{\begin{equation}\begin{aligned}}
\newcommand{\eea}{\end{aligned}\end{equation}}
\newcommand{\beq}{\begin{eqnarray}}
\newcommand{\eeq}{\end{eqnarray}}

\newcommand{\Li}{\ensuremath{\textrm{Li}}}

\newcommand{\hsigma}{\ensuremath{\hat{\sigma}}}
\newcommand{\rsigma}{\ensuremath{\rho}}
\newcommand{\sre}{\ensuremath{r^{*}_e}}
\newcommand{\stheta}{\ensuremath{\theta^{*}}}

\def\SU{SU}

\def\U{\mathrm{U}}

\DeclareMathOperator{\Tr}{Tr}

\DeclareMathOperator{\Vol}{Vol}

\begin{document}

\begin{titlepage}
\thispagestyle{empty}

\begin{flushright}
\normalsize
PUPT-2406
\end{flushright}
\vfil

\bigskip

\begin{center}
\LARGE Quivers, YBE and 3-manifolds
\end{center}

\vfil
\vspace{1.5cm}

\begin{center}
\def\thefootnote{\fnsymbol{footnote}}

Masahito Yamazaki

\bigskip
\vspace{0.3cm}

\small
Princeton Center for Theoretical Science,\\ Princeton University, Princeton, NJ 08544, USA

\end{center}

\vspace{1cm}

\begin{center}
{\bfseries Abstract}
\end{center}

\bigskip

We study 4d superconformal indices for a large class of $\scN=1$
superconformal quiver gauge theories realized combinatorially 
as a bipartite graph or a set of ``zig-zag paths'' on a two-dimensional torus $T^2$.
An exchange of loops, which we call a ``double Yang-Baxter move'', gives
 the Seiberg duality of 
the gauge theory, and the invariance of the index under the duality is 
translated into the Yang-Baxter-type equation
of a spin system defined on a ``Z-invariant'' lattice on $T^2$.
When we compactify the gauge theory to 3d, Higgs the theory and then
 compactify further to 2d,
the superconformal index reduces to an integral of quantum/classical dilogarithm
 functions.
The saddle point of this integral unexpectedly reproduces the hyperbolic volume of a
 hyperbolic 3-manifold. The 3-manifold is obtained by gluing hyperbolic
ideal polyhedra in $\bH^3$, each of which could be thought of as a 3d lift of
 the faces of the 2d bipartite graph.
The same quantity is also related with
the thermodynamic limit of the 
BPS partition function, or equivalently the
genus $0$ topological string partition function,
on a toric Calabi-Yau manifold dual to quiver gauge theories.
We also comment on brane realization of our theories.
This paper is a companion to another paper summarizing the results \cite{Terashima:2012cx}.

\vspace{3cm}

\end{titlepage}

\newpage

\setcounter{page}{1}

\section{Introduction and Summary}\label{sec.intro}

Four-dimensional supersymmetric quiver gauge theories
has been a useful playground to understand the
physics of strongly coupled phenomena of gauge theories,
in particular their IR fixed points.

In this paper, we study a large class of 4d $\scN=1$ quiver gauge
theories described combinatorially by a configuration of oriented cycles
(called {\it zig-zag paths}) on a
two-dimensional
torus, satisfying certain conditions analyzed below \cite{Hanany:2005ss}.
This combinatorial data, equivalently expressed
as a bipartite graph (dimer) or a quiver diagram on $T^2$,
 encodes
the matter content (the quiver diagram)
and the Lagrangian
of our gauge theories \cite{Hanany:2005ve,Franco:2005rj,Franco:2005sm}.
We here take the gauge group at each vertex of the quiver diagram to be $U(N)$.
The resulting gauge theory 
is believed to flow to a non-trivial interacting fixed point in the IR,
can be 
engineered from a stack of $N$ D3-branes
probing the tip of the a toric Calabi-Yau manifold,
and has been extensively studied in the context of AdS/CFT
correspondence (see \cite{Kennaway:2007tq,Yamazaki:2008bt} and
references therein).

There is an interesting subtlety in this story. The gauge
theory corresponding to a given toric Calabi-Yau manifold is not unique, 
and several different gauge theories, related by a sequence of 
Seiberg dualities \cite{Seiberg:1994pq}, correspond to the same geometry (this is sometimes
called ``toric duality'').
In the language of zig-zag paths, this is translated into an 
ambiguity of the relative position of the loops,
and Seiberg duality is translated into an exchange
of the loops, which we call a ``double Yang-Baxter move''.
As the naming suggests, this is the standard Yang-Baxter move repeated
twice, and strongly suggest the integrable structure behind the theory.

Given a 4d supersymmetric gauge theory,
we could extract concrete quantitative statements of the theory by 
computing its 4d superconformal index $I$ \cite{Romelsberger:2005eg,Kinney:2005ej}.
This is a twisted partition function on $S^3\times S^1$, where
the chemical potentials are turned on along the $S^1$ direction.
The index can be computed in the free field limit, 
and is written as a matrix integral.
One of the main results of this paper is that 
this matrix model could be regarded 
as the partition function of a
 spin system defined from zig-zag paths on $T^2$, where each spin 
has $N-1$ 
continuous values
in $S^1$. This is summarized in the relation
\beq
I_\textrm{4d quiver}=Z_\textrm{spin system on $T^2$} \ .
\label{I4d=Z2d}
\eeq
This is the manifestation of the integrable structure mentioned above;
the Seiberg duality is now translated into the statement
that the resulting partition function is invariant under the 
double Yang-Baxter move, ensuring the integrability of the model.
Interestingly, (modulo some important differences mentioned below)
the resulting spin system is 
essentially the same as the spin system studied in 
\cite{Bazhanov:2010kz} for $SU(2)$ gauge groups, and 
more recently in \cite{Bazhanov:2011mz} for $SU(N)$ gauge groups.

Let us next study the reduction of our 4d $\scN=1$ theories along
$S^1$. The resulting theory has 3d $\scN=2$ supersymmetry, and flows in
the IRg
to a non-trivial fixed point.
As the $S^1$ shrinks all the KK modes decouple, and the 4d
superconformal index should reduce to a partition function on $S^3$.
Indeed, it has been shown that the 4d index 
in this limit reduces (after suitably regularizing divergences)
to a 3d partition function on ellipsoid $S^3_b$
(defined in \eqref{S3b}), which could be again written as a matrix
integral after localization computation:
\beq
I_\textrm{4d quiver}\,  \longrightarrow \, Z_\textrm{3d on $S^3_b$} \ .
\eeq
This limit is also natural in the context of integrable models; the
Yang-Baxter equation, being an equality, should hold even after
taking the limit\footnote{As we will see there are some subtleties
associated with the regularization of divergences in the limit.}.
After taking one more limit explained in the text (Higgsing to the
Abelian gauge group),
the solution of the star-triangle relation studied in \cite{Bazhanov:2010kz}
reduces to another solution discovered by Faddeev and Volkov
\cite{Volkov:1992uv,FaddeevVolkovAbelian,FaddeevCurrent}, clarifying the
integrable structure behind 3d $\scN=2$ theories.

We could also consider further dimensional reduction to 2d. 
This is simply the $b\to 0$ limit of the ellipsoid partition function,
and taking the leading contribution we have 
\begin{align}
Z_{\textrm{3d on $S^3_b$}} \,  \longrightarrow \,  Z_{\textrm{2d on
$\bR^2$}} = \int  d\sigma\,  \exp\left[\frac{1}{2\pi b^2} \scW_{\textrm{2d}}(\sigma)+\scO(b^0)\right] \ ,
\end{align}
where $\sigma$ is the scalar component(s) of the twisted superfield
(defined from the derivative of the vector superfield) and 
takes values in the Cartan of the gauge group\footnote{In general $\sigma$ is a vector, but we here do not show this fact
explicitly for notational simplicity.}.
The potential $\scW_{\rm 2d}(\sigma)$ represents the effective twisted superpotential
obtained by integrating out matters from the theory.

The surprising observation, based on the works
\cite{BobenkoS,Bazhanov:2007mh}, is that this twisted effective
superpotential
is identified with the hyperbolic volume of a certain 3-manifold $M$,
in the case that $N=2$.


The 3-manifold $M$ is determined from the bipartite graph on $T^2$
which in turn is determined from zig-zag paths,
and could be thought of as a 2d graph with an ``extra dimension'' added.
The 3-manifold $M$ is defined as the 
union of ideal hyperbolic polyhedra in $\bH^3$,
and the projection of the polyhedra onto the 
boundary of $\bH^3$ gives the faces of the 2d bipartite graph. 

The twisted superfield scalars $\sigma$, in this description, is
identified with the geometric modulus of the 3-manifold $M$; the dihedral angles
of $M$ are determined from the radii of circles on $T^2$, whose
logarithms coincide with $\sigma$. The values of $\sigma$
are determined from the gluing conditions
of the 3-manifold $M$
\beq
\exp\left( \frac{\partial \textrm{Vol}[M](\sigma)}{\partial \sigma}\right)=1
\ . 
\label{Volextremize}
\eeq
There is a counterpart of this equation on the gauge theory side; the value of
$\sigma$ at the vacuum is determined from the equation
\beq
\exp\left(\frac{\partial \scW_{\rm 2d}}{\partial \sigma}\right)=1 \ .
\label{Wextremize}
\eeq
We find that the two conditions \eqref{Volextremize}, \eqref{Wextremize}
coincide. 
In other words, the vacua of the 2d $\scN=(2,2)$ theory is captured by
the gluing condition of the 3-manifold! This is the second main result
of our paper.


One quick supporting evidence for the correspondence between
\eqref{Volextremize} and \eqref{Wextremize} is that 
the twisted superpotential $\scW_{\rm 2d}$ is expressed as a sum
of the Lobachevsky functions (or classical dilogarithm functions),
and the same function is known to appear in the 
formula for the volume of hyperbolic tetrahedra;
$M$ is simply the sum of these tetrahedra.
Of course, the appearance of the dilogarithm function applies to any 
3d $\scN=2$ theories dimensionally reduced on $S^1$, whereas
our correspondence should hold only for a specific
class of 3d gauge theories.

In Table \ref{dict} we summarize our correspondence between 2d $\scN=(2,2)$ theory
and the geometry of the 3-manifold $M$. The data on both
sides come from the zig-zag paths on $T^2$, and therefore from a toric
Calabi-Yau 3-fold or from the brane configuration for our gauge theories. 

\begin{table}[htbp]
\begin{center}
\caption{Dictionary between the 2d $\scN=(2,2)$ theories and the 3-manifold.}
\medskip
\begin{tabular}{c|c}
   2d $\scN=(2,2)$ gauge theory & 3-manifold \\ 
\hline
\hline
  twisted superpotential $\scW_{\rm 2d}(\sigma)$  &
	 $\textrm{Vol}[M](\sigma)$ \\
\hline
  scalar in twisted superfield $\sigma$ & modulus $\sigma$ of $M$  \\  
\hline
  matter contributing $\textrm{Li}_2$ & tetrahedron contributing
	 $\textrm{Li}_2$ \\
\hline
  vacuum equation $\exp\left(\frac{\partial \scW}{\partial
     \sigma}\right)=1$ & gluing condition 
$\exp\left( \frac{\partial M}{\partial \sigma}\right)=1$
\end{tabular}
\label{dict}
\end{center}
\end{table}

The correspondence to this point refers only to the 2d gauge
theory. However, it is natural to ask if similar correspondence persists
for the 3d/4d gauge theories we started with. As for the 3d gauge
theory,
the natural guess is to
propose
\beq
Z_\textrm{3d on $S^3_b$} \sim  Z_\textrm{3d $SL(2)$ Chern-Simons on $M$} \ ,
\label{Z3dZ3d}
\eeq
where the right hand side is the (holomorphic) partition function of the 
$SL(2)$ Chern-Simons theory on $M$ and the parameter $b$ is identified
with the inverse square root of the level $t$ of the Chern-Simons
theory: $b^2\sim 1/(t+2)$. This is consistent with our previous correspondence 
since the classical limit of the $SL(2)$ Chern-Simons theory 
reproduces the volume (and the Chern-Simons invariant) of the 
3-manifold. There is also generalization of \eqref{Z3dZ3d}
to $N> 2$, where the right hand is replaced by the 
partition function of $SL(N)$ Chern-Simons theory on $M$.

The relation \eqref{Z3dZ3d}, relating 3d $\scN=2$ quiver gauge theories and the 3d
$SL(2)$ Chern-Simons theory, is highly reminiscent of the recently
found connection
\cite{Terashima:2011qi,Dimofte:2011jd,Terashima:2011xe,Dimofte:2011ju,Cecotti:2011iy,Nagao:2011aa,Dimofte:2011py} between 3d $\scN=2$ theories and 3d $SL(2)$
Chern-Simons theories (see also \cite{Dimofte:2010tz}). 
There one of the crucial
underlying data is the
Riemann surface and its Teichm\"{u}ller space,
whereas here we have a dimer as the crucial ingredient.
We expect that this similarity could be explained
from the equivalence of underlying mathematical structures
(for example, cluster algebras), perhaps along the lines of
\cite{Goncharov:2011hp} ({\it cf.} \cite{Franco:2011sz,Nagao:2011aa,Xie:2012dw}).
It would be interesting to understand the
precise relation between the two.

Let us also comment on the relation between \eqref{I4d=Z2d} and
\eqref{Z3dZ3d}.
It is natural to interpret both statements from compactification of a
6d theory (see the discussion in section \ref{subsec.brane}
and \ref{subsec.M5}). If this is true, then the two statements are
related by a dimensional reduction on the one hand, and by a dimensional
oxidation on the other side, thus exemplifying the statement
({\it cf.} \cite[section 5]{Benini:2011nc})
\begin{equation*}
\textrm{dimensional reduction = dimensional oxidation}
\end{equation*}
in the AGT\cite{Alday:2009aq}-type correspondence.

\bigskip

Finally, we point out connection of our results to topological string
theory and the BPS state counting.

Under an assumption about the bipartite graph
(isoradiality condition in section \ref{subsec.quivers}), we show that the critical value of the
hyperbolic volume of our 3-manifold $M$ could be written
as a sum of Lobachevsky functions (see \eqref{Volisoradial}).
Interestingly, exactly the same expression arises as a
Legendre transform of the
thermodynamic limit of the partition function of the dimer model.
This dimer model has been studied in the context of BPS state counting of 
type IIA string theory on a Calabi-Yau 3-manifold $X_{\Delta}$,
which in turn is known to be equivalent to the topological string partition function (modulo wall crossing phenomena).
In this context the thermodynamic limit is the semiclassical limit $g_{\rm top}\to 0$,
where $g_{\rm top}$ is the topological string coupling constant,
and the leading contribution is precisely the prepotential
$\scF_{\textrm{top}, 0}$ \cite{Ooguri:2009ri}.
Combining these observations, we have
\beq
 \scF_{\textrm{top}, 0}
 \textrm{ is an
integral of the Legendre
transformation of }
\textrm{Vol}[M_0] \ .
\label{VolLegendre}
\eeq

This paper is companion to \cite{Terashima:2012cx}, which announces basic results.

\begin{landscape}
\begin{figure}[htbp]
\centering{\includegraphics[scale=0.5]{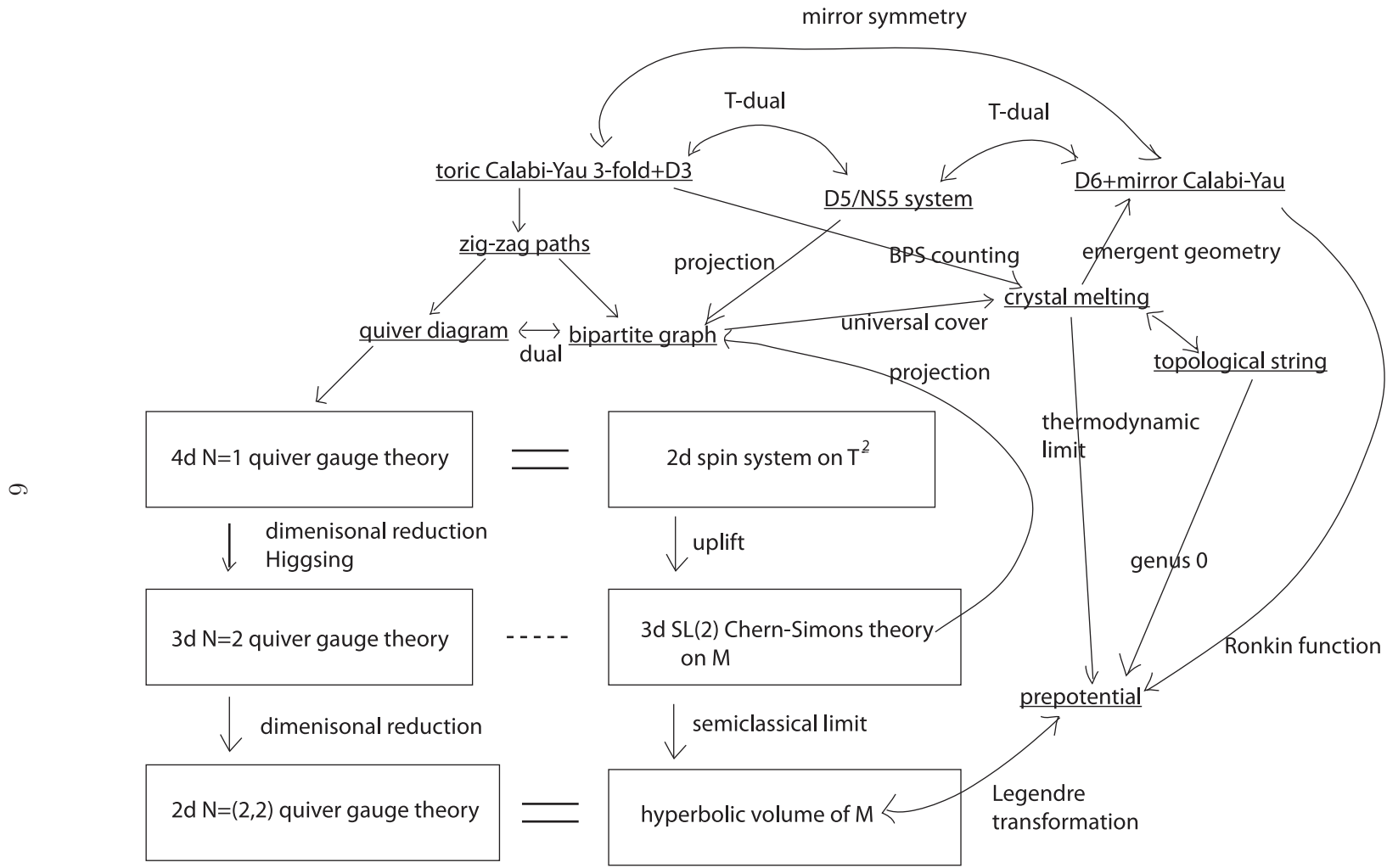}}
\caption{Logical structure of this paper. Clearly it is impossible to
 list all the connections between all the ingredients mentioned here.
The main claims of this paper are the two equalities represented in the center of this
 figure.}
\label{structure}
\end{figure}
\end{landscape}

This paper is organized as follows (see Figure \ref{structure} for the
logical structure of this paper).
After a summary of 4d $\scN=1$
quiver gauge theories described from zig-zag paths (section \ref{sec.quivers}),
we compute
its superconformal index and comment on the reformulation
as a spin system (section \ref{sec.indexasspin}). We then reduce the theory down to 
3d and 2d, and study the connection with 3-manifolds (section \ref{sec.reduction}). 
Section \ref{sec.dimers} explains the relation with topological string theory and 
the statistical mechanical model of BPS state melting.
We conclude with some future problems (section \ref{sec.conclusion}).
Appendix contains a summary of the special functions used in the main text,
and an explicit computation of thermodynamic limit of the 
dimer partition function.

\section{Quivers from Zig-Zag Paths} \label{sec.quivers}

In this section we briefly summarize 
the construction of 4d $\scN=1$ superconformal quiver theories 
from ``zig-zag paths'' on $T^2$ \cite{Hanany:2005ss,Gulotta:2008ef} (see
\cite{Ueda:2006jn,Goncharov:2011hp} for mathematical
formulation).
See also the reviews \cite{Kennaway:2007tq,Yamazaki:2008bt} for more
details on dimer model techniques.

\subsection{Zig-Zag Paths}\label{subsec.zigzag}

For the clarify of the presentation let us first explain the
combinatorial properties of the zig-zag paths,
which is actually rather elementary.
The physical context will be explained shortly.

Let us start with a convex polygon $\Delta$ 
in $\bZ^2$. Geometrically this is
the toric diagram for a Calabi-Yau 3-manifold $X_{\Delta}$,
{\it i.e.}, the cone $\Delta\times \{1\}\in \bR^3$
specifies the top-dimensional cone of the fan.

As a toric diagram there are $SL(3, \bZ)$ ambiguities in the choice of
$\Delta$.
For example, $\bZ^2$ translation of $\Delta$ keeps the geometry.
In the following we use the same symbol $\Delta$ for the equivalence class of
$\Delta$ under this identification.

One way to specify $\Delta$ is to write down 
the set of primitive normals of
the polygon. 
Let us denote them by $(r_i, s_i)\ne (0,0)$ with $i=1, \ldots, d$.
We choose the label $i$ such that the direction of the vector $p_i$ 
rotates in the counterclockwise
manner as we increase $i$.
The integer $d$, which is the number of lattice points in the boundary
of $\Delta$, is fixed throughout this paper.
Note that in general the same vector could appear multiple times in this
list. This happens when an edge at the boundary of $\Delta$ contains
 more than two lattice points.

By definition we have
\beq
\sum_i r_i=\sum_i s_i=0 \ .
\label{rssum}
\eeq

\bigskip

Let us now consider zig-zag paths.
The zig-zag paths are a set of closed oriented cycles
 $p_1, \ldots, p_d$ on a two-dimensional torus $T^2$, whose homologies
 cycles are determined by $(r_i, s_i)$: 
\beq
[p_i]=r_i[\alpha]+s_i[\beta] \in H_1(T^2, \bZ) \quad \textrm{fixed and non-trivial}  \ ,
\eeq
where $[\alpha], [\beta]$ are the basis of $H_1(T^2, \bZ)$, for
     example $\alpha$ and $\beta$-cycles of the torus.
There is $SL(2, \bZ)$ ambiguity in the choice of $[\alpha], [\beta]$,
which could be absorbed into the $SL(2, \bZ)$ ambiguity in the choice of
$\Delta$.
The origin of the terminology ``zig-zag path'' will be clarified shortly
when we study dimer models.

We assume the following three conditions.

\begin{figure}[htbp]
\centering{\includegraphics[scale=0.5]{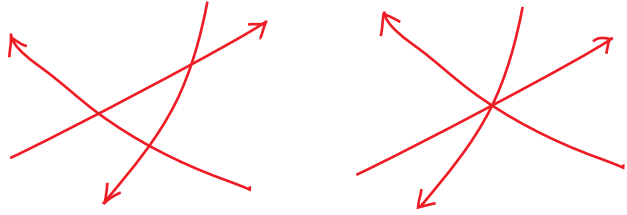}}
\caption{Genericity condition, stating that no three zig-zag paths
intersects at a single point. The left figure is allowed, whereas the right is not.}
\label{genericity}
\end{figure}

\begin{figure}[htbp]
\centering{\includegraphics[scale=0.5]{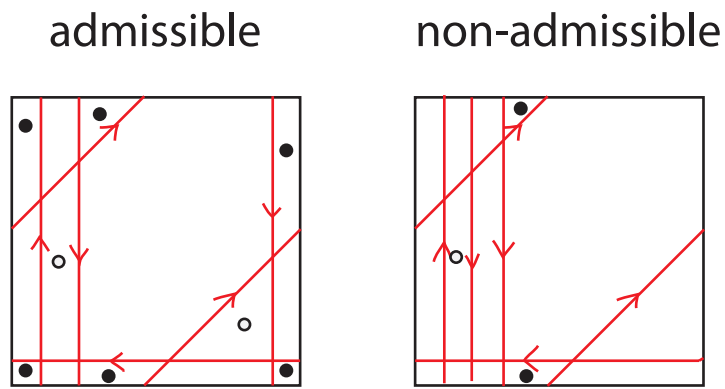}}
\caption{Admissible (left) and non-admissible (right) configuration of
 zig-zag paths. Rather than coloring the faces by black and white, we
 have represented the coloring by placing black and white dots inside
 (it is hard to represent the white color on a white paper!).
We see from this example that moving a zig-zag path across an intersection point 
breaks the admissibility condition.}
\label{admissibility}
\end{figure}

\begin{itemize}
\item
{\bf genericity.}
First, we assume that no three paths intersect at a single point
(Figure \ref{genericity}).
This is satisfied for generic choice of paths.

\item
{\bf admissibility.}
Second, we impose the {\it admissibility condition} (this terminology
     comes from \cite{Ueda:2006jn}). 
To explain this, let us note that the paths divide the torus into a union of convex polygons. We color
the convex polygon by black (white) if all the paths around the
polygon has counterclockwise (clockwise) orientation around the polygon;
otherwise the face is kept uncolored.
The paths are called admissible if every edge bounds a colored polygon
     (Figure \ref{admissibility}).
\end{itemize}

These two conditions are sufficient for the
4d quiver gauge theory. We moreover impose one simplifying assumption

\begin{figure}[htbp]
\centering{\includegraphics[scale=0.5]{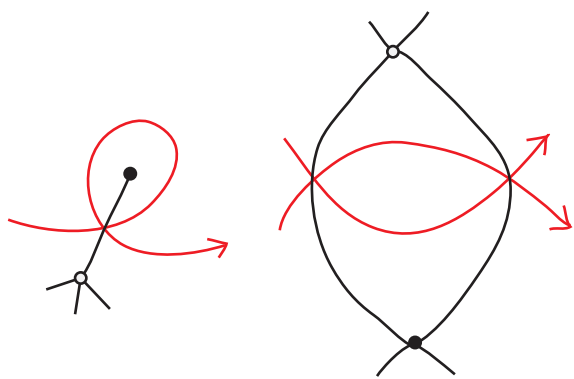}}
\caption{Minimality condition forbids two types of
 intersections of zig-zag paths. The graph
superimposed on it is the bipartite graph $\scG^*$.}
\label{minimality}
\end{figure}

\begin{itemize}
\item
{\bf minimality.}
Minimality (this terminology comes from \cite{Goncharov:2011hp}) forbids
     the
two possibilities shown in Figure \ref{minimality}.
\end{itemize}

We will come back to the physical significance of the admissibility
condition in section \ref{subsec.brane}, but for the moment 
let us first analyze its combinatorial implication by defining 
graphs on $T^2$.

Given a set of zig-zag paths 
we can define a natural bipartite
graph $\scG^*$ and its dual $\scG$, both realized on $T^2$
(the symbol is chosen for later
convenience). 
In the literature $\scG^*$ is often called
a brane tiling \cite{Hanany:2005ve, Franco:2005rj,Franco:2005sm},
and $\scG$ a periodic quiver.
The vertices of $\scG^*$ are given by colored
faces, and the edges by the intersection points between them. 
The orientation of zig-zag paths gives 
a natural orientation to the edges, 
and ensures that the resulting graph is
bipartite, {\it i.e}, vertices are colored either black or white and
edges connect vertices of different colors.

The graph $\scG^*$ is dual of $\scG$. 
This means the
vertices are placed on the uncolored faces, and the edges at each
intersection point between them. The orientation of the zig-zag paths
determine the orientation of the edges.
In the following we denote the set of vertices, edges and faces of $\scG$ by
$V, E$ and $F$. Since we have black and white colors
we have a decomposition into black and white faces:
$F=B\cup W$. 
For an edge $e\in E$ we denote the source (and the target) by $s(e)$
($t(e)$). Since
$\scG$
is drawn on $T^2$ we have
\beq
\big| V \big|-\big| E \big|+\big| F\big|=0 \ .
\label{VEF}
\eeq
We also denote by $V^*, E^*, F^*$ the set of vertices/edges/faces of the
graph $\scG^*$. By definition we have
$$
V^*=F, \quad E^*=E, \quad F^*=V \ .
$$

\begin{figure}[htbp]
\centering{\includegraphics[scale=0.5]{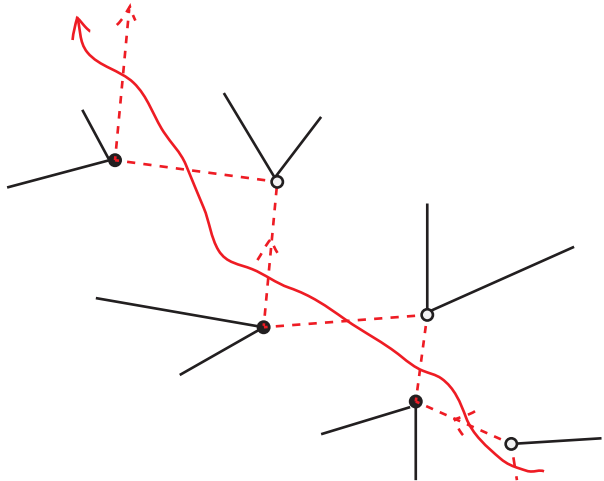}}
\caption{A zig-zag path on the bipartite graph $\scG^*$ (dotted path)
is identified with the zig-zag path defined previously (undotted arrow).}
\label{zigzagpath}
\end{figure}

We have defined $\scG, \scG^*$ from zig-zag paths, but we can go in the other
direction. Given a bipartite graph we define a zig-zag paths to be graph
on $\scG^*$ which turns maximally right (left) at black (white) vertex.
Because the graph is finite, we always come back to the same vertex after
several steps and hence this defines a set of closed loops. 
The name zig-zag path originates from the zig-zag shape of the 
path\footnote{This is also called a rhombus loop or a
train track in the literature. The word ``rhombus'' refers to a
quadrilateral in Figure \ref{trigonometrics}, which becomes a rhombus
for isoradial circle patterns.} (see Figure \ref{zigzagpath}).

\begin{figure}[htbp]
\centering{\includegraphics[scale=0.45]{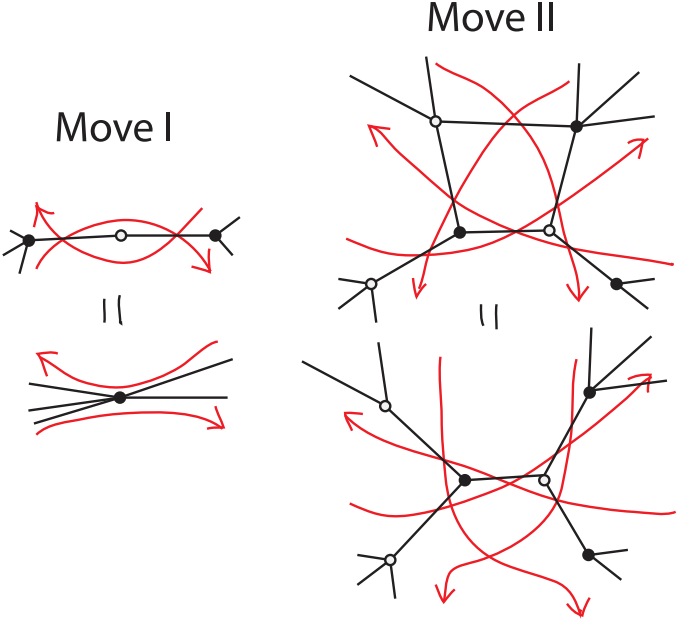}}
\caption{Two basic moves (fundamental moves) 
preserving the admissibility condition. The
 second move is called a double Yang-Baxter move.}
\label{moves}
\end{figure}

We can verify that this gives an inverse to our previous construction,
{\it i.e.}, from the bipartite graph we recover the zig-zag paths we
started with. However,
it is important to
notice that this correspondence is not one-to-one;
we can start with an
admissible configuration to obtain another configuration by locally
applying the two basic moves shown in Figure \ref{moves}.
These two moves are called {\it fundamental moves} in this paper
and complete in the sense that any two minimal 
bipartite graphs corresponding to
the same $\Delta$ are related by a sequence of these
two moves \cite[Theorem 2.5]{Goncharov:2011hp}. The same theorem
guarantees the existence of minimal admissible configuration of zig-zag paths.
Note that the number of nodes of the quiver
is preserved in the fundamental moves.

\begin{figure}[htbp]
\centering{\includegraphics[scale=0.45]{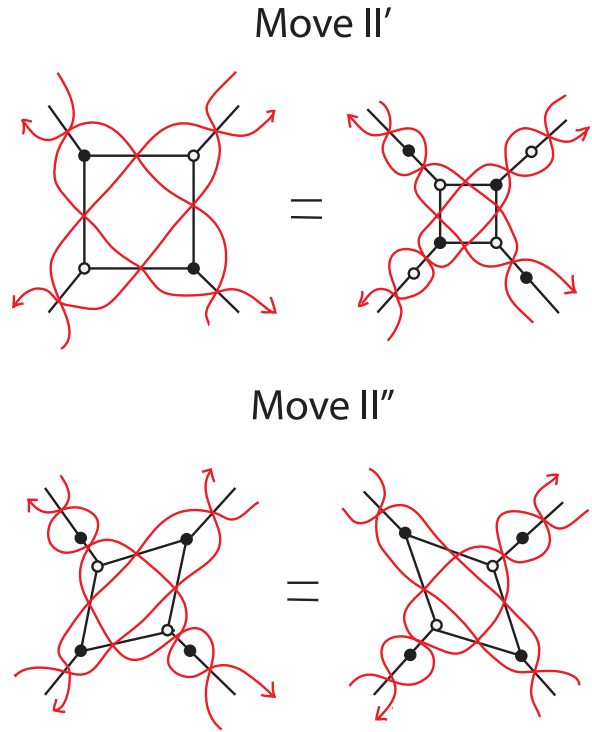}}
\caption{We can replace the fundamental move II in Figure \ref{moves}
by either of the two moves shown here.}
\label{moves2}
\end{figure}

\begin{figure}[htbp]
\centering{\includegraphics[scale=0.5]{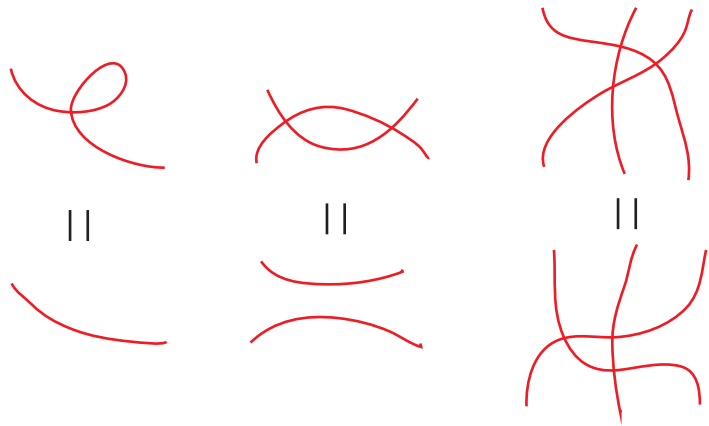}}
\caption{Moves for zig-zag paths without admissibility condition. Here
 we do not show the orientation of the zig-zag paths, and all
 orientations are allowed.}
\label{moveswo}
\end{figure}

It should be kept in mind that the choice of the fundamental moves is not
unique.
For example, we could replace the second move in Figure \ref{moves2}
by a different move, for example the ones shown in Figure \ref{moves}.
We can easily check that the these moves, in combination with move I,
generate the same set of moves.
We call move II' a Seiberg move (it is exactly Seiberg duality,
as we will see shortly)
and move II'' a spider move.
For later reference, we also list basic moves for zig-zag paths 
without admissibility condition imposed (Figure \ref{moveswo}). 
These are reminiscent of the Reidemeister moves of knot theory.

\subsection{Quiver Gauge Theories}\label{subsec.quivers}

Let us next define 4d $\scN=1$ quiver gauge theories from the combinatorial data
of the previous subsection. This is simply a quiver gauge theory
determined from $\scG$. In other words, we have a gauge group
$SU(N)_v$ for each vertex $v\in V$, and a bifundamental chiral multiplet
$X_e$ for each 
$e\in E$.
The total gauge group is given by
\beq
G=\prod_{v\in V }SU(N)_v \ .
\eeq
Here and in the following we will specialize to the case where the ranks of the gauge groups
are all equal to $N$.
Most of the analysis of the next section on 4d superconformal index
generalizes straightforwardly to more general cases where the ranks are
position dependent. However, the analysis of the Seiberg duality in 4d,
and also of the reduction to 3d, requires some change.

We also determine the superpotential to be
\beq
W=\sum_{b\in B} \textrm{Tr}\left(\prod_{e\in b}X_e\right)-
\sum_{w\in W}\textrm{Tr}\left(\prod_{e\in w}X_e\right) \ ,
\label{superpotential}
\eeq
where the product inside the trace is taken in the counterclockwise
(clockwise) manner for $B$ and $W$.

The quiver gauge theory constructed in this way contains examples with
enhanced supersymmetries, for example the $\scN=4$ theory or $\scN=2$ theories
corresponding to the $A_L$ type singularities \cite{Douglas:1996sw}.
However, generically the theory is chiral, and has $\scN=1$
supersymmetry. The typical example in the literature is the theory dual to the conifold \cite{Klebanov:1998hh}.

Although our theory is in general chiral, there is no chiral anomaly.
This is because
by construction the number of incoming and outgoing arrows for each
vertex $v\in V$ are the same.

The quiver gauge theories constructed in this way is the world-volume
effective theory on the D3-branes probing the
toric Calabi-Yau manifold $X_{\Delta}$, and flows in the IR to a
non-trivial IR fixed point.

The basic moves in Figure \ref{moves}
is an operation on quiver gauge theories which keeps the IR fixed point
intact \cite{Franco:2005rj}.
The first move is to remove two fields $X, Y$ with superpotential $\Tr(XY)$
from the theory --- because this superpotential term represents a mass
term, we can simply integrate out a massive field.
The second move (move II') corresponds to a Seiberg duality (see
\cite{Yamazaki:2008bt}, section 4.7 for detailed exposition),
or equivalently the mutation of the quiver.

Note that in our setup Seiberg duality can be taken only for those nodes
which has four arrows (two incoming and two outgoing).
This is because $N_f$ is always a multiplet of $N$, and the only value
of $N_f$ in the conformal window is $N_f=2 N$\footnote{
It is important that not all the possible mutations of the quiver physically
make sense. If we mutate on the $n$-valent vertex with $n\ge 6$, 
the result is
in general a graph which could not be realized on torus.}.

Finally, let us comment on a further simplifying assumption on the bipartite graph.
A bipartite graph is called isoradial if all the vertices can be placed
on circles of equal radius. In terms of zig-zag paths
this is satisfied if and only if 
(1) each zig-zag path is a simple closed curve
and (2) the lift of any pair of zig-zag paths to the universal cover intersect at most once \cite{KenyonSchlenker}.
Colloquially this means that all the zig-zag paths are ``straight enough''
such that the zig-zag paths have minimal intersection numbers.
This means that the number of edge of the quiver is minimal, {\it i.e.,}
we have 
\beq
| E |=\sum_{i<j} |\langle p_i, p_j \rangle | =\sum_{i<j} |r_i s_j-r_j s_i | \ .
\eeq
We will see that this condition leads to enormous simplifications of
part of the upcoming
analysis; for example, in section \ref{subsec.2d} we will comment on this
condition in the context of 2d gauge theories.
However, it should be kept in mind that isoradiality is not a necessary
condition, and for example stronger than the consistency conditions in
\cite{Gulotta:2008ef,UedaIshiiConsistency,Mozgovoy:2008fd}.

\subsection{Brane Configuration}\label{subsec.brane}

The quiver gauge theories in the previous subsection could be realized
by D5/NS5 brane configurations \cite{Imamura:2006ie,Imamura:2007dc,Yamazaki:2008bt}. 
Let us briefly summarize this, since 
this clarifies the origin of the apparently ad hoc assumptions in the
previous subsection.

The relevant brane configuration is 
\beq
\begin{tabular}{c||c|c|c|c|c|c|c|c|c|c}
 &  0 & 1& 2 & 3& 4& 5& 6& 7& 8& 9 \\
\hline
$N$ D5 &  -- & -- & -- & --   & &  $\cdot$ &  &  $\cdot$ & &  \\
\hline
NS5 &  --& --& -- & -- & \multicolumn{4}{c|}{$\Sigma$} & &  
\end{tabular}
\eeq
Here $N$ D5-branes wraps the $T^2$ along the 57
directions,
whereas a single NS5-brane wraps a holomorphic cycle $\Sigma$
in 4567 direction.
If we write $x=e^{x_4+i x_5}, y=e^{x_6+i x_7}$, then the holomorphic
curve is given by
\beq
\Sigma=\{ P(x, y)=0 \} \subset (\bC^{\times})^2  \ , 
\label{mirrorcurve}
\eeq
where $P$ the so-called Newton polynomial of $\Delta$, defined by
\beq
P(x,y)=\sum_{(i,j)\in \Delta} c_{i,j} x^i y^j \ .
\eeq
with generic coefficients $c_{i,j}$. 
Here $(i,j)\in \Delta$ means the lattice point $(i,j)$ is
a lattice point of $\Delta$ (including the boundary).
The curve $\Sigma$ is a Riemann surface of genus $g$ and puncture $d$,
where $g$ is the number of internal lattice points inside $\Delta$ (recall that $d$ also denotes the number of zig-zag paths in section \ref{subsec.zigzag}).
This curve is part of the mirror $\check{X}_{\Delta}$
of the toric Calabi-Yau 3-fold $X_{\Delta}$ given by \cite{Hori:2000ck,Hori:2000kt}
\beq
\check{X}_{\Delta}: uv+ P(x,y)=0 \ ,
\label{mirrorCY3}
\eeq
with $u, v\in \bC$.
We will encounter $P(x, y)$
again as a spectral curve of the dimer model.
The identification of the NS5-brane curve and the mirror curve
could be explained by a T-duality, where our D5/NS5 system is 
mapped to a configuration of D6-branes wrapping Lagrangian 3-cycles
inside $\check{X}_{\Delta}$ \cite{Feng:2005gw}.

This brane configuration can be looked at from two different
ways: D5-brane viewpoint and the NS5-brane viewpoint.
In the former, we have a $T^2$, which is the torus we had previously,
and the zig-zag paths represent the intersection cycles of an NS5-brane
and the D5-brane; and the colored faces represents the projection of the 
shape of the NS5-brane. 
In this language,
the colored regions correspond to the projection of the mirror curve
into the 57 directions (called coamoeba/alga \cite{Feng:2005gw}),
and the black/white color represents the orientation of the curve when
projected onto $T^2$.
In other words, black/white regions represents 
$(N, 1)$/$(N, -1)$-branes,
and uncolored regions $(N,0)$-branes.
This explains why $U(N)$ gauge group lives in uncolored regions.
The bifundamental fields originate from the massless strings between the $U(N)$
gauge groups, namely the intersection points of the uncolored regions.
The admissibility condition simply says we do not have $(N, k)$-branes
with $|k|\ge 2$, in which case no Lagrangian descriptions are known.

We could also take the NS5-brane viewpoint. Then we have a Riemann
surface $\Sigma$, on which we have a set of 1-cycles representing 
the intersection with D5-branes. 
Our Riemann surface $\Sigma$ is
reconstructed as the surface whose boundaries are the cycles of 
zig-zag paths \cite{Feng:2005gw}.
This is parallel to the construction of the Seifert surface  
in knot theory, and 
Seiberg duality in the original graph is translated into the 
half Dehn twist of $\Sigma$ \cite{Goncharov:2011hp}. 
We will encounter this Riemann surface $\Sigma$ again in section \ref{subsec.M5}.

\subsection{R-charges}\label{subsec.R-charge}

We are going to consider the IR fixed point of our quiver gauge
theories. Due to the strong coupling effects the bifundamental chiralmultiplets could have large anomalous
dimensions in the IR. 
This is determined from the IR $U(1)_R$ R-symmetry in the superconformal algebra.
In general, it is a rather non-trivial problem to identify the IR superconformal
R-symmetry $U(1)_R$, because UV R-symmetry mixes with the global UV $U(1)$
symmetries.
Here we will comment on one particular useful parametrization 
of the UV global symmetry or equivalently IR R-symmetry (see
\cite{Hanany:2005ss,Imamura:2006ie,Imamura:2007dc}\footnote{The paper
\cite{Hanany:2005ss} discuss the case of isoradial bipartite graphs, however 
our parametrization applies to more general bipartite graphs. This will
be crucial when we discuss Seiberg duality in section \ref{subsec.moves}.}).

We are going to define $d$ global symmetries, with $1$ relation among them.
For each zig-zag path $p_i$, let us define the charge of the 
bifundamental field $X_e$ at an edge $e \in E$ by
\beq
Q_i[X_e]=\langle p_i, e\rangle \ ,
\label{paramrule}
\eeq
where the braket here refers to the (signed) intersection number of the two 
paths $p_i$ and $e$.
To see that this is in fact a global symmetry, recall that 
a term in the superpotential is represented by a closed loop 
around a black/white vertex of the bipartite graph (see \eqref{superpotential}),
and has $0$ intersection number  with a closed loop $p_i$.
Because two zig-zag paths pass through $e$ with an opposite orientation,
we find that the diagonal subgroup of these $d$ global symmetries is trivial
(see \ref{rssum})
\beq
\sum_i Q_i[X_e] = 0 \ . 
\eeq
Hence we find $d-1$ global symmetries. These symmetries are anomaly free
\cite{Butti:2006hc}, because 
\beq
\sum_{e:\, \textrm{around }v} Q_i[X_e] = 0 \quad \textrm{for all } i \ .
\label{Qianomalyfree}
\eeq
This holds because a zig-zag path incoming to a vertex
necessarily goes out of the vertex, with an opposite orientation
and hence with a field with an opposite flavor charge.

It is known that generically this parametrization exhausts all the possible 
global symmetries\footnote{In some speical cases there could be an enhancement to non-Abelian
global symmetries.}. There is a corresponding statement in the AdS dual; 
two out of $d-1$ correspond to isometries of the Sasaki-Einstein manifold
and called mesonic symmetries, whereas the remaining $d-3$ symmetries are associated with the 3-cycle of the Sasaki-Einstein manifold
and called baryonic symmetries.

We can also describe R-symmetries. The choice of UV R-symmetry is not
unique, since
we can consider a mixing with the $d-1$
global symmetries mentioned above (the IR R-symmetry inside superconformal
algebra is determined by a-maximization \cite{Intriligator:2003jj}).

\begin{figure}[htbp]
\centering{\includegraphics[scale=0.5]{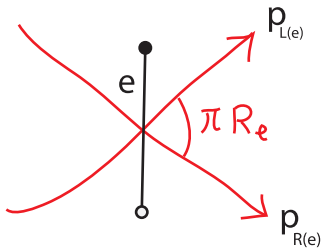}}
\caption{An bifundamental at an edge $e$ has flavor charge $+1$ for
 $L(e)$-th flavor charge and $-1$ for $R(e)$-th flavor charge, where
 $p_{L(e)}, p_{R(e)}$
are two zig-zag paths as in this Figure.}
\label{flavorbifund}
\end{figure}

For our purpose, a particularly useful parametrization is given as
follows. Let us choose a set of $d$ parameters $\theta_i$ for each zig-zag path
$p_i$, and let us assume that they are defined modulo $2\pi$ 
($\theta_i\sim \theta_i+2\pi$),
and that $0\le \theta_{i+1}-\theta_i\le \pi$ 
for all $i$ ($\theta_{d+1}:=\theta_1$).
We can regard $\theta_i$ as the slope of $p_i$.
Then for each bifundamental $X_e$ for $e\in E$ its R-charge is 
simply defined to be the relative slopes of the two 
zig-zag paths which goes through the edge.
More formally, we define 
\beq
R_e=R[X_e]:=\frac{1}{\pi}\textrm{sign}\langle p_{L(e)}, p_{R(e)}\rangle [\theta_{L(e)}-\theta_{R(e)}] \ ,
\label{Rchargerule}
\eeq
where $[x]$ denotes a real number in $[0,2\pi]$ and equivalent to $x$
modulo $2\pi$. 
We also used the notation that a chiralmultiplet corresponding to an edge $e$
has a flavor charge $+1$ for the $L(e)$-th zig-zag path and $-1$ for the
$R(e)$-th path, where $p_{L(e)}$ and $p_{R(e)}$ are two zig-zag paths passing
through $e$ 
from opposite sides (see Figure \ref{flavorbifund}). 
From the definition we have $R_e\ge 0$.

We have the following two conditions on IR R-symmetries.
\begin{itemize}
\item
First, the $\beta$-function for Yukawa
     couplings vanish.
This is the same as the requirement that
the  R-charge of the superpotential, and therefore any term in the
     superpotential, is normalized to be 2. This means
\beq
\sum_{e\in F} R_e=2 \ .
\label{sum}
\eeq

\item
Second, the $\beta$-functions for the gauge coupling vanish. 
From the NSVZ $\beta$-function, which in our case could be written as
\beq
\frac{d}{d\log \mu}\frac{1}{g_v^2}=
\frac{N}{1-g_v^2 N/8\pi^2}\left[
3-\frac{1}{2}\sum_{e\in v}(1-\gamma_e)
\right] \ ,
\label{gaugebeta}
\eeq
where the anomalous dimension $\gamma_e$ is related to the R-charge
by $\gamma_e=3R_e-2$.
From this condition, we have
\beq
\sum_{e\in V} (1-R_e)=2 \ .
\label{sumdual}
\eeq
This can be written more symmetrically 
\beq
\sum_{e\in F^*} R_e^*=2 \ , 
\label{sumdual2}
\eeq
where we defined $R_e^*:=1-R_e$.
\end{itemize}
These two conditions follow from the definition
\eqref{Rchargerule} and the fact that the sum of exterior angles of a
polygon is $2\pi$.

Naively the dimension of the solution space to \eqref{sum},
\eqref{sumdual} is zero because we have $|E|$ parameters $R_e$ and
$|V|+|F|$ constraints (see \eqref{VEF}).
However, in supersymmetric theories not all the constraints are
independent
and it has been demonstrated that the solution has $d-1$ parameters \cite{Imamura:2007dc}.
Again, it is straightforward to see that the two conditions \eqref{sum},
\eqref{sumdual2} are preserved under the deformation with the $d-1$
parameters. Of course, this is the same as the number of global
symmetries treated above.

\section{Superconformal Index As Spin System}\label{sec.indexasspin}

In this section we first define the superconformal index for 4d $\scN=1$
superconformal gauge theories. We then show that the 4d index 
for the spin system defined in previous section is 
equivalent to the partition function of a spin system on $T^2$.

\subsection{Superconformal Index}\label{subsec.index}

\noindent\underline{The Definition} 

Let us consider 4d $\scN=1$ superconformal theory on $S^1\times S^3$.
This theory has supercharges $\scQ_{\alpha}, \overline{\scQ}_{\dot{\alpha}}$
and $\scS_{\alpha}, \overline{\scS}_{\dot{\alpha}}$,
where $\alpha$ and $\dot{\alpha}$ denotes the
spins $\SU(2)_1$ and $\SU(2)_2$ 
of the isometry of $S^3$: $Spin(4)=\SU(2)_1 \times \SU(2)_2$. 
To define an index, we need to pick up a particular supercharge.
There are four supercharges, but $\scQ_+$ and $\scQ_-$
($\overline{\scQ}_{\dot{+}}$
and $\overline{\scQ}_{\dot{-}}$) define the same index due to the 
$\SU(1)_1$ ($\SU(2)_2$) symmetry, and hence we have two possibilities.
When we choose $\scQ=\scQ_{-}$, the superconformal index is defined as an index defined from
$\scQ$, with insertions of operators commuting with $\scQ$:
\beq
I^L(t,y;z)=\textrm{Tr}\left[ (-1)^F t^{2(\scE+j_2)} y^{2j_1} u^\scF
e^{-\gamma \{\scQ, \scQ^{\dagger} \} } \right]\ ,
\label{ILdef1}
\eeq
where the index is taken over the Hilbert space on $S^3$.
$F$ is the fermion number, $\scE$ is the energy, $j_1$ and $j_2$ are the spins of $SU(2)_1$ and
$SU(2)_2$, respectively, and 
$u^{\scF}:=\prod_i u_i^{\scF_i}$, where $\scF_i$ ($u_i$) is the charge 
(chemical potential) under the
$i$-th flavor 
symmetry\footnote{This is often denoted by $z_i$ in the literature.}.
Note that due to the translation symmetry we could flip the sign of
$j_1$,
and we have
\beq
I^L(t,y;u)=I^L(t,y^{-1};u) \ .
\label{flipI}
\eeq
This index, often called the left-handed index (hence the symbol $L$ in
\eqref{ILdef1}) is independent of the value of $\gamma$ thanks to the
standard index argument,
and can be computed in the limit $\gamma\to \infty$
\beq
I^L(t,y;u)=\textrm{Tr}\left[  (-1)^F t^{2(\scE+j_2)} y^{2j_1} u^\scF \right]\  ,
\eeq
where the trace is now taken over all states satisfying 
\beq
\{ \scQ, \scQ^{\dagger}\}=\scE-2j_2+\frac{3}{2}r=0 \ ,
\eeq
where $r$ is the $U(1)$ R-charge.
Similarly, if we choose
$\overline{\scQ}=\overline{\scQ}_{\dot{-}}$, we can define the right-handed index
\beq
I^R(t,y;u)=\textrm{Tr}\left[(-1)^F t^{2(E+j_1)} y^{2j_2} u^F \right]\  ,
\eeq
where the trace is taken over all states satisfying
\beq
\{ \overline{\scQ}, \overline{\scQ}^{\dagger}\}=E-2j_1-\frac{3}{2}r=0 \ .
\eeq
This is the same as the left-handed index, except that the 
orientation of the arrows of the quiver diagrams are reversed.
Since this can be taken into account by a change of convention,
in the following we will concentrate on the left-handed index
and denote the corresponding index simply by $I$.

For our later purposes it is useful to reparametrize the chemical
potentials as
\beq
p=t^3 y, \quad q=t^3 y^{-1} \ .
\eeq
In this notation, the superconformal index reads
\beq
I^L(p,q;u)=\textrm{Tr}\left[ (-1)^F p^{\frac{E+j_2}{3}+j_1} q^{\frac{E+j_2}{3}-j_1}   u^F e^{-\gamma \{ Q, Q^{\dagger} \} }\right] \  .
\eeq

One subtlety we have is that our 4d quiver theory
is defined in UV, and is conformal only in
the IR, where the theory is strongly coupled.
Because the index is independent of the continuous parameter,
it is independent of the dimensionless parameter 
obtained by multiplying the energy scale by the radius of $S^3$,
and we could compute the
index in UV, 
except that we have to take into account the
mixing of UV R-symmetry with global symmetry.
In our analysis this is taken into account when we have included
chemical potentials for global symmetries. In practice this means that the effect of anomalous dimensional could be taken into effect by shifting the global symmetry chemical potentials by powers of $t$.
\bigskip

\noindent\underline{Integral Expression} 

Because an index is invariant under the continuous deformation of parameters
of the Lagrangian,
the superconformal index can be computed by taking the free field
limit. Alternatively, we could apply localization techniques.
In either way, the result is written as an integral
over the Cartan of the 
Cartan $H$ of the gauge group, which in our case is given by
\beq
H=\left( \U(1)^{N-1} \right)^{|V|} \subset G= \SU(N)^{|V|} \ .
\eeq
We parametrize an element of the Cartan of $SU(N)_v$ (at vertex $v\in V$)
by $z_v=(z_{v,1}, \ldots, z_{v,N})$ satisfying $\prod_{i=1}^N z_{v,,i}=1$.
We also write $z_{v,i}=e^{i \sigma_{v, i}}$, where
$\sigma_{v, i}$ is periodic with period $2\pi$.
Physically these parameters represents the Polyakov loop along
the thermal direction $S^1$.

The index is expressed in a plethystics form (this follows from group
theory, see \cite{Kinney:2005ej})
\beq
I(p,q,u)=\int \prod_{v\in V} \left[ d\mu_v \prod_{i<j} (z_{v,i}-z_{v,j})(z_{v,i}^{-1}-z_{v,j}^{-1}) \right]\,
\label{IndexIntegral}
\exp\left(
\sum_{n=0}^{\infty} i(p^n,q^n,u^n;z^n) 
\right) \ .
\label{Iassum}
\eeq
Here the integration measure contains $d\mu_v$ defined by
\beq
d\mu_v=\frac{1}{N!} \prod_{i=1}^{N-1} \frac{dz_v}{2\pi i z_v} \ ,
\eeq
as well as the Vandermonde determinant, 
and the integral over the contour $|z_{v, i}|=1$ (or $\sigma_{v, i}$
runs from $0$ to $2\pi$).
The ``single-letter index'' $i(p,q,u;z)$ is given as a sum over
contributions from vector and chiralmultiplets
\beq
i(p,q,u;z)
=\sum_{v\in V}  i^v_{\rm vect}(p,q; z)
+
\sum_{e\in E} i^e_{\rm chiral}(p,q,u;z) \ ,
\eeq
where 
\begin{align}
& i^v_{\rm vect}(p,q;z)=\left[1-\frac{1-pq}{(1-p)(1-q)}\right] \chi_{\rm adj}(z_v), \\
& i^e_{\rm chiral}(p,q,u;z)=\frac{1}{(1-p)(1-q)}
\left[
(pq)^{\frac{R_e}{2} }y u_{L(e)} u_{R(e)}^{-1}
\chi_{\rm bifund}(z_{s(e)},z'_{t(e)})  \right. \nonumber \\
& \hspace{5.5cm}
 - \left.
(pq)^{1-\frac{R_e}{2}}/(yu_{L(e)} u_{R(e)}^{-1}) \chi_{\rm
 bifund}(z_{s(e)}^{-1}, z_{t(e)}^{'\,-1})
\right]
\ , 
%
%
\end{align}
and $\chi_{\rm adj}$ and $\chi_{\rm bifund}$ are the characters
for the adjoint and bifundamental representations, respectively
\beq
\chi_{\rm adj}(z)=\sum_{1\le i, j\le N}(z_iz_j^{-1})-1,
\quad
\chi_{\rm bifund}(z,z')=
\sum_{i,j=1}^N 
z_i z'{}_j^{-1} \ .
\eeq
Note that vector multiplets are not charged under flavor symmetries.

For our purposes, it is useful to rewrite the index in a different form
\beq
I(p,q,u)=\int   \prod_{v\in V} d\mu_v
\, %
\left( \prod_{v\in V}  I^v_{\rm vect}(p,q;z)\right) 
\left( \prod_{e\in E} I^e_{\rm chiral}(p,q,u;z) \right) \ ,
\label{Iasproduct}
\eeq
where 
\beq
I^v_{\rm vect}(p,q;z)=
\kappa(p,q)^{N-1}
\prod_{k\ne
l}\frac{1}{\Gamma(z_{v,k} z_{v, l}^{-1};p,q)} \ , 
\eeq
and
\beq
I^e_{\rm chiral}(p,q;z)=\prod_{1\le k, l \le N}
\Gamma( (pq)^{\frac{R_e}{2}} u_{L(e)}
 u_{R(e)}^{-1} z_{s(e), k} z_{t(e), l}^{-1};p,q) \ .
\label{IhyperN}
\eeq
Here $\Gamma(z;p,q)$ is the elliptic gamma function 
defined in \eqref{ellGammadef}
and we defined (see \eqref{infbracket})
\beq
\kappa(p,q):=(p;p)_{\infty}(q;q)_{\infty} \ .
\eeq

The equivalence of the two expressions \eqref{Iassum} and
\eqref{Iasproduct}
can be verified by the equations of the form
\beq
\exp\left(\sum_{n=1}^{\infty}
\frac{1}{1-x^n}\right)=\prod_{m=0}^{\infty}
\frac{1}{1-x^m} \ .
\eeq
and the equalities in Appendix, for example \eqref{usefulgamma}.

We here obtained the expression of the index \eqref{Iasproduct}
by rewriting
\eqref{Iassum}.
However it should be emphasized that \eqref{Iasproduct} arises
directly in the localization derivations of the index 
(see for example \cite{Imamura:2011uw} and \cite[Appendix]{Benini:2011nc}).
The factors $I^v_{\rm vect}$ ($I^e_{\rm matter}$) represents the 1-loop
determinants for vector multiplet at vertex $v$ (chiralmultiplet at edge
$e$), and the infinite product comes from the spherical harmonics
expansion on $S^3$. In the free field computation, we have the $S^3$
Laplacian
from the bosons, whose determinant could be written as products over the
quantum numbers of the spherical harmonics, and the same applies to
fermions. There are cancellations between bosons and fermions, and the
unpaired bosonic (fermionic) modes appear in the denominator (numerator)
of the product \eqref{ellGammadef}.

\subsection{Invariance under Fundamental Moves}\label{subsec.moves}

In this section we prove an invariance the superconformal
index under the fundamental moves. 
The general argument that the 4d index depends 
only at the IR fixed point guarantees this invariance, but 
we can also check this explicitly.

\bigskip
\noindent\underline{Generalities on Gluing}

It is important to note that the two fundamental moves in Figure \ref{moves} are local operations on the quiver diagram. We therefore expect that 
the invariance of the index should reduce to the invariance 
of the index defined for the subdiagram.

To formalize this idea we need to 
invoke the concept of ``gluing'' in gauge theories.
Let us first explain this in a rather general situation\footnote{This
discussion obviously generalizes to quantum field theories 
in other dimensions, for example to ellipsoid partition 
function in 3d studied in section \ref{subsec.3d}.}.
Consider two 4d $\scN=1$ theories $T_1$ and $T_2$
which has global symmetries $G_1$ and $G_2$.
Suppose moreover that these flow in the IR to non-trivial fixed points.
We can then compute the superconformal indices for each of these
theories,
with the chemical potentials for global symmetries included.

\begin{figure}[htbp]
\centering{\includegraphics[scale=0.35]{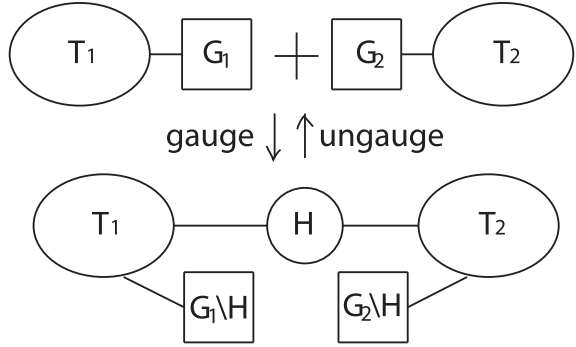}}
\caption{We can glue two theories $T_1$ and $T_2$ by gauging a common
 global symmetry $H$.}
\label{gluing}
\end{figure}

Now from the two global symmetries $G_1$ and $G_2$
we choose a common subgroup $H$ ($H_1\subset G_1$, $H_2\subset G_2$,
$H\simeq H_1\simeq H_2$)
and gauge $H$, {\it i.e.,} the diagonal subgroup of $H_1\times H_2$. 
The resulting theory has global symmetry $G_1\backslash H_1\times
G_2\backslash H_2$, where $G\backslash H$ denotes the commutant of $H$
inside $G$,
see Figure \ref{gluing}.

This gluing operation has a counterpart at the level of the index.
Let us denote the indices of $T_1$ and $T_2$ by $I_1(p,q;u_1, w)$ and $I_2(p,q;u_2, w)$, where $u_1$ and $u_2$ denotes the chemical potentials for the Cartan of the flavor symmetries $G_1\backslash H_1$ and $G_2\backslash H_2$, respectively, and $w$ for those for $H$.
We then have\footnote{This is the gluing of ``generalized index'', see \cite{Kapustin:2011jm} for
similar analysis in 3d.
}
\beq
I(p,q;u_1, u_2)=\int \frac{dw}{2\pi i w}
\int I_1(p,q;u_1,w) \, I_{\rm vect}(w) \, I_2(p,q;w,u_2) \  .
\label{Idecomp}
\eeq
Note that after gluing the global symmetry is promoted to
dynamical degrees of freedom to be integrated over, 
and hence we have included the 1-loop determinant for $w$.

\bigskip
Let us apply this to our case, where one of our moves
 replaces a quiver diagram with another.
Since the fundamental moves are local operations on the graph, we could 
decompose the quiver diagram  $\Gamma$
into two parts along edges of $\Gamma$ such that
$\Gamma_1$ becomes $\Gamma_2$ after the move
and $\Gamma\backslash \Gamma_1=\Gamma\backslash\Gamma_2$ 
kept intact.
Then the rule in \eqref{Idecomp} immediately gives
\beq
I_{\Gamma}=\int d\mu(z) \, I_{\rm vect}(p,q)\, I_{\Gamma_1}(p,q;z)
\, I_{\Gamma/\Gamma_1}(p,q;z) \ ,
\eeq
and
\beq
I_{\Gamma'}=\int d\mu(z) \,  I_{\rm vect}(p,q)\, I_{\Gamma_2}(p,q;z)\, I_{\Gamma/\Gamma_2}(p,q;z)
\ .
\eeq
The equality of these two quantities follow if we could show the local invariance of the index 
\beq
I_{\Gamma_1}=I_{\Gamma_2} \ .
\eeq
Because any duality is 
generated by combination of the two moves (section \ref{subsec.zigzag}), 
all we need to do is to check the invariance under the two fundamental moves.

\bigskip
\noindent\underline{Integrating Out}

Let us first study move I.
After the move we have two extra bifundamental fields
$X_1$ and $X_2$ between two gauge group $SU(N)_v$ and $SU(N)_{v'}$, which has a superpotential term $\textrm{Tr} (X_1 X_2)$.
This means that the sum of R-charges is two and their global symmetries
charges have opposite signs, and the contribution from the two is
\beq
\Gamma\left( (pq)^{R_e/2} z_v z_{v'}^{-1} u_i
u_j^{-1};p,q\right)\Gamma\left((pq)^{(2-R_e)/2} z_{v'} z_v^{-1} u_i^{-1} u_j
;p,q\right) \ .
\label{integrateout}
\eeq
But this is trivial due to \eqref{Gammainverse}.

\bigskip
\noindent\underline{Seiberg Duality}

Let us first analyze move II', which is equivalent to the Seiberg duality.
The invariance of the index under Seiberg duality has been verified in
\cite{Dolan:2008qi}. However, an extra analysis is required here because
the invariance of the index holds only for particular assignment of
R-charges to fields, and 
we need to check that the assignment of the R-charge in section
\ref{subsec.R-charge} satisfies this condition.

\begin{figure}[htbp]
\centering{\includegraphics[scale=0.5]{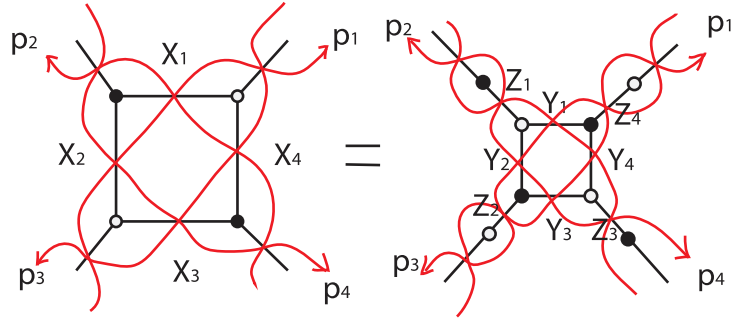}}
\caption{The labeling scheme for the bifundamental fields and
 zig-zag paths for Seiberg dual theories.}
\label{SeibergR}
\end{figure}

As we demonstrated already we can concentrate on the part of the bipartite
graph
which changes under the duality.
Let us label the fields of the electric theory
by $X_i$ and the magnetic theory by $Y_i, Z_i$ ($i=1, \ldots, 4$)
(see Figure \ref{SeibergR}). Following the rule \eqref{Rchargerule}
and using the quantity
\beq
R_{ij}:=\frac{1}{\pi}\textrm{sign}\langle p_i, p_j\rangle 
\left[\theta_i-\theta_j\right] \ ,
\eeq
we can parametrize the R-charges as 
\beq
 R[X_1]=R_{12}, \quad
 R[X_2]=R_{41}, \quad
 R[X_3]=R_{34}, \quad
 R[X_4]=R_{23} \ ,
\eeq
in the electric theory and 
\begin{equation}
\begin{split}
 &R[Y_1]=R_{34}, \quad
 R[Y_2]=R_{23}, \quad
 R[Y_3]=R_{12}, \quad
R[Y_4]=R_{41} \ ,
\\  
 &R[Z_1]=R_{42}, \quad
 R[Z_2]=R_{31}, \quad
 R[Z_3]=R_{24}, \quad
R[Z_4]=R_{13} \ .
\end{split}
\end{equation}
in the magnetic theory.
Note that there are relations
\beq
R[X_1]=R[Y_3], \quad R[X_2]=R[Y_4], \quad R[X_3]=R[Y_1], \quad
R[X_4]=R[Y_2] \ ,
\label{Rcond1}
\eeq
and 
\begin{equation}
\begin{split}
R[Z_1]&=R[X_1]+R[X_2], \quad R[Z_2]=R[X_2]+R[X_3], \\
R[Z_3]&=R[X_3]+R[X_4], \quad R[Z_4]=R[X_4]+R[X_1] \  .
\label{Rcond2}
\end{split}
\end{equation} 
The last four equations are natural since $Z_i$ is the meson composed 
of two electric quarks $X_i$'s.

Let us compute the superconformal index.
The standard Seiberg duality claims that electric and magnetic theories flow to the same IR fixed point.
In the electric theory we have $SU(N)$ gauge theory with $N_f=2N$
flavors.
This theory has in general
$SU(N_f)\times SU(N_f)$ flavor symmetries, but in our case we choose
$SU(N)^4$ subgroup and denote the 
chemical potentials by $(s_{1,i}, s_{2,i})$ for fundamental flavors
and $(t_{1,i}, t_{2,i})$ for antifundamental flavors ($i=1, \ldots, N$),
satisfying $\prod_i s_{k,i }=\prod_i t_{k,i}=1$ for $k=1, 2$.
From the rule of the previous subsection the index is
\begin{equation}
\begin{split}
I_E(p,q)=& \frac{\kappa(p,q)^{N-1}}{N!}
\int \prod_{i=1}^{N-1} \frac{dz_j}{2\pi i z_j}
 \displaystyle\frac{
\prod_{i=1}^N
\prod_{a=1}^{2N} 
\Gamma(S_a z_j ;p,q) \Gamma(T_a z_i^{-1}; p,q)}{\prod_{i\ne j}\Gamma(z_i
z_j^{-1}; p, q)} \ ,
\end{split}
\end{equation}
for
\begin{equation}
\begin{split}
\{ S_a\} &=\{ (pq)^{\frac{R[X_1]}{2}}s_{1,i} u_1 u_2^{-1}, \quad
(pq)^{\frac{R[X_3]}{2}}s_{2,i} u_3 u_4^{-1}
\}  \ ,
\\
\{ T_a\} &=\{ (pq)^{\frac{R[X_2]}{2}}t_{1,i}^{-1} u_4 u_1^{-1}, \quad
(pq)^{\frac{R[X_4]}{2}}t_{2,i}^{-1} u_2 u_3^{-1} \} \ .
\end{split}
\end{equation}
The magnetic theory is again has $SU(N)$ gauge group with $2N$
flavors, and the index contains contributions from mesons:
\begin{equation}
\begin{split}
I_M=\frac{\kappa(p,q)^{N-1}}{N!}
& 
\prod_{a,b=1}^{2N}\Gamma(\tilde{U}_{a,b};p,q)
\int \prod_{i=1}^{N-1} \frac{dz_j}{2\pi i z_j}
 \displaystyle\frac{
\prod_{i=1}^N
\prod_{a=1}^{2N} 
\Gamma(\tilde{S}_a z_j ;p,q) \Gamma(\tilde{T}_a z_i^{-1}; p,q)}{\prod_{i\ne j}\Gamma(z_i
z_j^{-1}; p, q)} \ ,
\end{split}
\end{equation}
for
\begin{equation}
\begin{split}
\{ \tilde{S}_a\} &=\{(pq)^{\frac{R[Y_1]}{2}}s_{1,i}^{-1} u_3 u_4^{-1},
\quad  (pq)^{\frac{R[Y_3]}{2}}s_{2,i}^{-1} u_1 u_2^{-1}
\}  \ ,
\\
\{ \tilde{T}_a\} &=\{(pq)^{\frac{R[Y_2]}{2}}t_{1,i} u_2 u_3^{-1}, \quad
 (pq)^{\frac{R[Y_4]}{2}}t_{2,i} u_4 u_1^{-1} 
\} \ .
\end{split}
\end{equation}
and
\begin{equation}
\begin{split}
\{ \tilde{U}_{a,b} \}=\{ (pq)^{\frac{R[Z_1]}{2}} s_{1,i}^{-1}t_{1,i} u_4
u_2^{-1}, 
 (pq)^{\frac{R[Z_2]}{2}} s_{2,i}^{-1}t_{1,i} u_3 u_1^{-1},  \\
 (pq)^{\frac{R[Z_3]}{2}} s_{2,i}^{-1} t_{2,i} u_2 u_4^{-1}, 
 (pq)^{\frac{R[Z_4]}{2}} s_{1,i}^{-1} t_{2,i} u_1 u_3^{-1}
\} \ .
\end{split}
\end{equation}
What we want to prove is the equivalence of the two expressions
\beq
I_E(p,q)=I_M(p,q) \ .
\label{IEM}
\eeq
As in \cite{Dolan:2008qi}, we will establish this with 
the help of a remarkable
identity of elliptic hypergeometric functions proven by
\cite{RainsTransf}, which is one of the many identifies studied, for
example, in \cite{Spiridonov:2009za,Spiridonov:2011hf}.

In order to apply \cite{RainsTransf}, we need to check the 
``balancing condition''. To state this, define
\begin{equation}
\begin{split}
S:&=(\prod_a S_a)^{1/N}=(pq)^{\frac{R[X_1]+R[X_3]}{2}}  u_1 u_2^{-1} u_3
u_4^{-1}, \\\
T:&=(\prod_a T_a)^{1/N}=(pq)^{\frac{R[X_2]+R[X_4]}{2}}  u_1^{-1} u_2  u_3^{-1}
u_4 \ .
\end{split}
\end{equation}
The balancing condition states that 
\beq
ST=pq \ ,
\eeq
which follows from the expression for $S$ and $T$ above.
In this case we have $I_E=I_M$, provided
\beq
\tilde{S}_a=S/S_a, \quad \tilde{T}_a=T/T_a, \quad
\tilde{U}_{a,b}=S_a T_b \ .
\label{balancing}
\eeq
These conditions follow from the equalities \eqref{Rcond1}, \eqref{Rcond2}.

It should also be kept in mind that 
to establish the identify above \eqref{IEM}
we do not need to know the 
exact values of the R-charges determined from a-maximization;
the equality of the index holds before a-maximization.

\subsection{Z-invariant Spin System} \label{subsec.spin}

We next show that the superconformal index for our quiver gauge theory
can be 
reformulated as a 
classical spin model on a lattice in $T^2$.

The basic idea is simple. 
Let us first regard the integral variables $\sigma^v=(\sigma^v_1, \ldots, \sigma^v_N)$
as a $N$-component continuous spin variable at vertex $v\in V$
(recall that these parameters are related to $z_v$ by $z_v=e^{i \sigma_v}$).
These variables are circle valued with period $2\pi$,
and in the $SU(N)$ case the $N$ components satisfy a constraint
$\sum_{i=1}^N \sigma^v_i=0$.

Next we need the Boltzmann weights associated with spin configurations.
This is determined from the 1-loop determinant
$I^v_{\rm vect}$ and $I^e_{\rm chiral}$:
\beq
I_{\rm 4d}=Z_{\rm spin}=\int \left(\prod_e d\sigma_e\right) e^{-\sum_{e\in E}
\mathcal{E}_e[\sigma]-
\sum_{v\in V} \mathcal{E}_v[\sigma]} \ ,
\eeq
where we introduced a new expression
\beq
e^{-\mathcal{E}_v}=I^v_{\rm vector}, \quad e^{-\mathcal{E}_e}=I^e_{\rm
chiral} \ .
\eeq
In this language, $I^v_{\rm vect}$ is regarded as the self-intersection of the spins at position $v\in V$,
and the latter, $I^e_{\rm chiral}$, is the nearest-neighbor interaction of the spins at positions
$s(e), t(e)\in V$.

To some readers this might look like a trivial re-naming of
what we already know. However, first note that
it is not true for general quiver gauge theories that resulting spin system is 
realized on $T^2$; the assumption of toric Calabi-Yau dual
was crucial for this fact.
Second, what is surprising about this spin system is that 
it is integrable.
One simple way to see this is that the invariance of the index under the move II is the invariance of the partition function of the spin system 
under the double Yang-Baxter move, and the invariance under the Yang-Baxter move is one form
of the integrability of the model. The chemical potentials 
of the 4d index ($p, q, u$) are regarded as the rapidity variables.
In fact, the spin system constructed above coincides with the
spin system studied by Bazhanov and Sergeev (hereafter BS) \cite{Bazhanov:2010kz,Bazhanov:2011mz}, modulo
some differences mentioned below.

One technical difference is that the BS model are defined on the plane $\bR^2$, whereas our
model is defined on $T^2$.
Another more essential difference is that the integrable models
have an invariance under a single Yang-Baxter move, whereas our index
has an invariance only under the double Yang-Baxter move.

Let us here explain this difference in more detail.
In the BS model, to realize a single Yang-Baxter move we have to 
abandon the admissibility condition on zig-zag paths.
Even in this case we could still choose a checkerboard pattern for the
faces,
corresponding to the colored and uncolored faces in our previous
discussion.
More concretely, in the brane realization explained in section \ref{subsec.brane} regions with $(N, k)$-brane with $k$ even ($k$ odd) are colored (uncolored).
We associate the spin variables $\sigma_{v, i}$ to uncolored regions,
where $i=1, \ldots, N$ and $v$ is the label for the uncolored region.

\begin{figure}[htbp]
\centering{\includegraphics[scale=0.5]{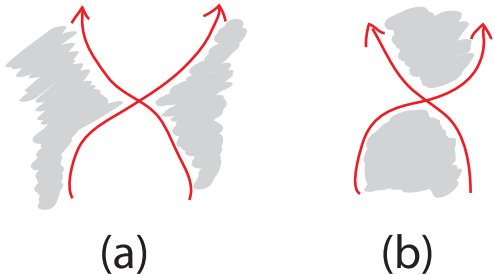}}
\caption{In admissible configuration of zig-zag paths we only have 
(a) as a possibility. However, if we lift admissibility condition, 
there are two possible
 types of intersections,
and correspondingly we need two different weights.
Red lines represents zig-zag paths, and gray region represents colored region.}
\label{weights}
\end{figure}

To define the Boltzmann weight, first we define the self energy 
$e^{-\scE^{\rm BS}_v[\sigma]}$ as before.
The definition of the nearest neighbor interaction $e^{-\scE^{\rm
BS}_e[\sigma]}$ is more tricky,
since in this general case we have two different
types of edges, shown in Figure \ref{weights}.
For the two
possibilities
BS model assigns two different weights,
which was denoted by $\scW$ and $\overline{\scW}$.

What BS has shown is that for a judicious choice of $\scE_v^{\rm BS}$
and $\scE^{\rm BS}_{e}$, the model is integrable
and is invariant under moves shown in Figure \ref{moveswo},
including the single Yang-Baxter move.
For $N=2$
the weight 
satisfies the star-triangle relation,
and this reduces to Spiridonov's formula for the elliptic
hypergeometric function \cite{Spiridonov:2010em}.
For $N>2$, 
the start-star relation is
still a conjecture, although there is non-trivial evidence
from 
power series expansion, see \cite{Bazhanov:2010kz}.

How is this related to our index?
First, let us specialize the BS model to the case with admissible zig-zag paths.
Then the case (b) in Figure \ref{weights}
does not arise. The claim is that in this case the Boltzmann weights
of BS model coincides with that of the 4d index, up to 
a spin-independent (but rapidity-dependent) 
overall normalization of the partition function.

Let us see this explicitly for $N=2$. In this case 
we can write $z_v=(z_v, z_v^{-1})$
and the weights are
\beq
I^e_{\rm chiral}(z_{s(e)}, z_{t(e)})=\Gamma((pq)^{1/2} z_{s(e)}^{\pm 1} z_{t(e)}^{\pm
1} u_{L(e)} u_{R(e)}^{-1};p,q) \
, 
\eeq
where here we meant the product of four terms with plus/minus signs.
We also have (see \eqref{Gammainverse} and
\eqref{thetarelation})\footnote{Contrary to \cite{Bazhanov:2010kz}
we have included the factor $1/N!$
in the measure, not in $I_{\rm vect}$.}
\beq
I^v_{\rm vect}(z)=
(pq)^{-1/8} \theta_1\left(\frac{\sigma_v}{\pi} \Big|\, p^{1/2}\right) \theta_1\left(\frac{\sigma_v}{\pi} \Big|\, q^{1/2}\right)  \ .
\eeq
These are identified with the weights function $\scW_{\alpha}$ and
$\scS$ of \cite{Bazhanov:2010kz}\footnote{The parameters $p, q$ here are
denoted by $p^2, q^2$ in \cite{Bazhanov:2010kz}.}, except by an overall spin-independent normalization factor for
$I_{\rm chiral}$.

The subtle difference in the normalization
arises since statistical physicists and 
gauge theorists have different motivations.
First, BS model requires a stronger 
invariance, {\it i.e.}, invariance under
moves violating admissibility conditions (Figure \ref{moveswo}), 
and the specific normalization
was chosen to serve that purpose. This is not necessary for 4d index.
Second, 4d index has a very specific normalization build into the
definition, whereas as a statistical mechanical problem
we can multiply $I_{\rm vector}$ by a constant depending on the chemical
potentials $p, q$ and $u$. For example,
the two moves of Figure \ref{moves} preserves the number of the
vertices of the quiver, and 
any overall multiplicative factor associated
with the vertex hence 
preserves
the invariance under the two moves.
In fact, 
contrary to \cite{Bazhanov:2010kz}, we are not normalizing the values of
the partition function in the thermodynamic limit.

As we explained in section \ref{subsec.brane},
the reason we stick to the admissibility condition is that 
we do not have a known Lagrangian for brane configurations for general
D5/NS5 brane configurations.
The fact that we can extend the integrable spin system to non-admissible
configuration
could be suggesting that 
the Seiberg duality extends to 
theories on multiplet NS5-branes.

\bigskip
\noindent\underline{Comments}

Several comments are now in order. 

First, the statistical mechanical model here is invariant under the 
moves in Figure \ref{moveswo}, and has Z-invariance
\cite{Baxter:1978xr,Baxter:1986df} (studied originally by Baxter in the context of 
eight-vertex models).

Second, in general the Boltzmann weight for the model
is not real and positive \cite{Bazhanov:2011mz},
which might look problematic 
as a statistical mechanical model.
However, in our context we are computing the index
and the answer necessarily comes with negative contributions.

Third, note that the resulting spin model is in general chiral,
except for the case $N=2$, where there is no distinction between
fundamental and anti-fundamental representation.
This explains why the multi-component spin models of
\cite{Bazhanov:2011mz} requires the chirality (and therefore extra
combinatorial data), as opposed to the 
$N=2$ case of \cite{Bazhanov:2010kz}.

Fourth, in the integrable models there is a duality transformation exchanging 
the graph $\scG$ with its dual $\scG^{*}$, which is a generalization of
duality of the Using model exchanging high-temperature expansion with
the low-temperature expansion. In our gauge theory context this duality
is broken; for example $\scG^*$ is a bipartite graph, but $\scG$ is not in
general.

Fifth, in the analysis to this point we have concentrated on the 4d index on
$S^1\times S^3$. However, as is clear from the argument above 
the specific choice of this observable in
itself does not matter --- other quantities will serve the same purpose 
as long as (1) they are invariant under the Seiberg duality
and integrating out massive matters (2) the quantity takes a factorized form as in \eqref{IndexIntegral}.

As an example of such quantity, we could use the 
4d lens space index computed in \cite{Benini:2011nc}.
This is a index on $S^1\times L(p,q)$, where $L(p, q)$,
with $p$ and $q$ coprime integers, is the lens space
($\bZ_p$ orbifold of $S^3$).
The complication for this case is that we have a set of integers, which
parametrize the discrete Wilson line. This means that the corresponding
spin chain has integer as well as continuous labels, and 
could not be interpreted as a conventional statistical mechanical model.
This lens space index has a reduction to 3d superconformal index in the $p\to \infty$ limit \cite{Benini:2011nc}, which suggests a generalization of the contents of the next section concerning 3d $S^3$ partition function to the 3d superconformal index.


Finally, let us make a brief comment on the comparison with the existing literature.
The invariance of the index under the Seiberg duality
for quiver gauge theories has been studied for some examples in \cite{Gadde:2010en}.
The results of this section apply to 
more general quiver gauge theories dual to an
arbitrary toric Calabi-Yau manifold.
Note also the relation of the 4d index and the statistical mechanical model
is known to experts, for example in \cite{Spiridonov:2010em}.
However, to the best of my knowledge the correspondence with the
supersymmetric 
quiver gauge theories has never been worked out in detail, and as we have seen there are in fact some subtle differences between the statistical mechanical models in the literature and the spin system defined from the 4d superconformal index.
Moreover, as we will see in the next section we will see that this insight leads to a remarkable
connection between 3d $\scN=2$ gauge theories and the geometry of 3d hyperbolic 3-manifolds.

\section{Reduction to 3d and 2d}\label{sec.reduction}

\subsection{Reduction to 3d}\label{subsec.3d}

The spin system defined 
in the previous section is based on a
rather general solution of the Yang-Baxter-type equation,
and 
several known integrable models, including the Kashiwara-Miwa model \cite{Kashiwara:1986tu} and
the
chiral Potts model \cite{Baxter:1987eq,AuYang:1987zc,vonGehlen:1984bi}
(and their $sl_N$ generalizations), arise
as a specialization/limit of the solution.
Therefore it is natural to ask whether there are some natural limits
which is of direct interest in the context of gauge
theory.

In gauge theory there is one obvious limit. Since we have a 4d theory on
$S^1\times S^3$, we could dimensionally reduce along the $S^1$.
The resulting theory has 3d $\scN=2$ symmetry, and by flowing to the IR 
we obtain a new 3d SCFT. We always need to keep in mind that there is
non-trivial RG flow involved in this process; in particular, the
anomalous dimensions in 4d and in 3d are in general different.
This reduction is also a natural in terms of integrable models --- it
is
simply the high-temperature limit of the theory.

The effect of this $S^1$ reduction on the superconformal index has been
studied in \cite{Dolan:2011rp,Gadde:2011ia,Imamura:2011uw,Nishioka:2011dq}, and we obtain a 3d
partition function on ellipsoid $S^3_b$ \cite{Hama:2011ea} (generalizing
the earlier results for $S^3_{b=1}$ \cite{Kapustin:2009kz,Jafferis:2010un,Hama:2010av})
\beq
S^3_b=\big\{ (z_1, z_2) \in \bC^2 \big|\,  b^2|z_1|^2+b^{-2}|z_2|^2=1 \big\} \ .
\label{S3b}
\eeq

To see this explicitly, first note that 
when the thermal $S^1$ shrinks all the chemical potentials $p, q, u$ go
to $1$, but we could keep the ratio fixed and finite:\footnote{
In $t, y, u$ variables this is to take
\beq
t=e^{-\beta/3}, \quad \eta=e^{-\beta\eta}, \quad u_i=e^{-\beta \mu_i} \ , 
\eeq
which coincides with the limit taken in \cite{Gadde:2011ia} for 4d $\scN=2$
theories.}
\beq
p=e^{-\beta (1+\eta)}, \quad q=e^{-\beta (1-\eta)}, \quad u_i=e^{-\beta \mu_i} \ .
\eeq
For a 1-loop determinant of the chiralmultiplet we have (see \eqref{IhyperN})
\beq
I^e_{\rm chiral}\! =\prod_{j,k} \prod_{m, n \ge 0}
 \frac{1-e^{-\beta \left[
  - i (\sigma_{s(e), j}-\sigma_{t(e), k} ) -(R_e-1+\mu_{L(e)}-\mu_{R(e)})+ (m+\frac{1}{2}) (1+\eta)
 +(n+\frac{1}{2}) (1-\eta)
\right]
 }}
 {
 1-e^{-\beta \left[
   i (\sigma_{s(e), j}-\sigma_{t(e), k} ) +(R_e-1+\mu_{L(e)}-\mu_{R(e)})+ (m+\frac{1}{2}) (1+\eta) +(n+\frac{1}{2}) (1-\eta)
\right]
 }
 } 
 \ .
\eeq
In the limit $\beta\to 0$ we regularize the expression
$1-e^{-\beta x}$ to be
\beq
[x]_{\beta}:= \frac{1-e^{-\beta x}}{1-e^{-\beta}}\to x, \quad \textrm{as } \beta\to 0 \ , 
\eeq
and $I^e_{\rm chiral}$ reduce to $Z^e_{\rm chiral}$, which is given by
\begin{align}
Z^e_{\rm chiral} &=\prod_{j, k =1}^N \prod_{m, n\ge 0}
 \frac{
  - i (\sigma_{s(e),j}-\sigma_{t(e),k} ) -(r_e-1)+ (m+\frac{1}{2}) (1+\eta)
 +(n+\frac{1}{2}) (1-\eta)
 }
 {
    i (\sigma_{s(e),j}-\sigma_{t(e),k} ) +(r_e-1)+ (m+\frac{1}{2}) (1+\eta)
 +(n+\frac{1}{2}) (1-\eta)
 }
\nonumber
\\
&=\prod_{j, k =1}^N \prod_{m, n\ge 0}
 \frac{
  \frac{Q}{2}\left(- i (\sigma_{s(e),j}-\sigma_{t(e),k} ) -(r_e-1)\right)+ (m+\frac{1}{2}) b
 +(n+\frac{1}{2}) b^{-1}
 }
 {
  \frac{Q}{2}\left( i (\sigma_{s(e),j}-\sigma_{t(e),k} ) +(r_e-1)\right)+ (m+\frac{1}{2}) b
 +(n+\frac{1}{2}) b^{-1}
 } 
\nonumber
\\
&=\prod_{j,k=1}^N
 s_b\left(\hat{\sigma}_{s(e), j}-\hat{\sigma}_{t(e), k}+\frac{iQ}{2}\left(1-r_e\right)\right)
 \ ,
\label{Z3dhyper}
\end{align}
where we defined
\beq
b=\sqrt{\frac{1+\eta}{1-\eta}}\ ,
\eeq
and $s_b(z)$, the quantum dilogarithm function defined in appendix, 
is used to regularize the infinite product \eqref{sbpole}.
In the last line we re-defined $\sigma$ by a factor of $\frac{Q}{2}$: $\hat{\sigma}:=\frac{Q}{2}
\sigma$.

We also defined the R-charge $r_e$ by
\beq
r_e=R_e+\mu_{L(e)}-\mu_{R(e)} \ .
\label{RIR}
\eeq
Here the parameters $\mu_i$ play the role of the real mass parameters obtained by weakly 
gauging the $i$-th global symmetry. 
The equation \eqref{RIR}
shows that the real mass parameters has the effect of changing
the anomalous dimension ({\rm cf.} \cite{Festuccia:2011ws}).
This is consistent with the R\"omelsberger's prescription
explained in section \ref{subsec.index}.

Similar analysis shows (after we regularize the divergence coming from
$\kappa(p, q)$)
\begin{align}
I^v_{\rm vect} \to Z^v_{\rm vect}
&=\prod_{i<j}
s_b\left(\hsigma_{v,i}-\hsigma_{v,j}+\frac{iQ}{2}\right)s_b\left(-(\hsigma_{v,i}-\hat{\sigma}_{v,j})+\frac{iQ}{2}\right)
\nonumber   \ , \\
&=\prod_{i<j }
4 \sinh b \left( \hsigma_{v,i}-\hsigma_{v,j} \right) \sinh b^{-1} 
\left( \hsigma_{v,i}-\hsigma_{v,j} \right) \ . 
\end{align}
We will drop the overall multiplicative constant $4$ in the following.
Note that 
the invariance of the index under $y\leftrightarrow y^{-1}$
\eqref{flipI}
is translated into the 
invariance $b\leftrightarrow b^{-1}$.
The full partition is still written as a matrix integral
of these 1-loop determinants
\beq
Z_{\rm 3d}=\int \left(\prod_{v\in V}d\sigma_v\right)
\left(\prod_{v\in V} Z^v_{\rm vector}\right)
\left(\prod_{e\in E} Z^e_{\rm chiral} \right) \ ,
\label{Z3d}
\eeq
which is precisely the 3d partition function on an ellipsoid.
Note that in our story not all the real mass parameters of
 the theory are turned on.
This is because we started from the 4d theory;
some 3d global symmetries are anomalous in 4d
and cannot be included in the 4d index.

Our 4d superconformal index has an invariance with respect to the two
fundamental moves. Our 3d partition function to have the same
property, 
since the latter is simply the limit of the former.
For example, 
in the same notation as in \eqref{integrateout},
the effect of move I is represented by
\beq
\prod_{i, j}
s_b\left(\hat{\sigma}_{v, i}-\hat{\sigma}_{v',
j}+\frac{iQ}{2}(1-r_e)\right)
\prod_{i, j}
s_b\left(\hat{\sigma}_{v', i}-\hat{\sigma}_{v, j}
+\frac{iQ}{2}(1-(2-r_e))\right) \ .
\eeq
This is trivial due to the relation \eqref{sbinverse}. Of course,
this is simply the limit of the relation \eqref{integrateout}.

\subsection{Higgsing}\label{subsec.Abelianization}

To make contact with the geometry of hyperbolic 3-manifolds,
we need to do one more reduction.
For clarify let us specialize to the case $N=2$ from here on.

As stated in the introduction, we give a VEV to the
vector multiplet scalar for the diagonal 
gauge group $U(1)_{\rm diag}$
\beq
U(1)_{\rm diag} \subset SU(2)^{|V|} \ ,
\eeq
and send the VEV to infinity.

More concretely, 
let us write the vector multiplet scalar for the $U(1)_{\rm diag}$ 
symmetry as $\hsigma$, and 
Higgs the $U(1)_{\rm diag}$ gauge symmetry
by sending $\sigma$ to infinity:
\beq
\hat{\sigma}^{\rm old}_v\to \hat{\sigma}^{\rm new}_v+\hat{\sigma}, 
\quad \sum_{v\in V} \hat{\sigma}^{\rm new}_v=0, \quad \hat{\sigma}\to \infty  \ .
\label{reduction}
\eeq
In this limit, the vector multiplet 1-loop determinant diverges to a constant
independent of $\hsigma_v$, and after regularization we have
\beq
Z^v_{\rm vector}\to Z'{}^v_{\rm vector}=1 \ . 
\eeq
This is simply the 1-loop determinant for the Abelian gauge theory, 
which means that the gauge group after Higgsing is simply given by
\beq
\left[\prod_{v\in V } U(1)_v\right] \, \Big/\, U(1)_{\rm diag}\ .
\eeq
For the chiralmultiplet, the 1-loop determinant in \eqref{Z3dhyper}
for $N=2$ reads
\beq
Z^e_{\rm
chiral}=
\frac{s_b\left(\hsigma_{s(e)}+\hsigma_{t(e)}+
\frac{iQ}{2}\sre\right) 
s_b(\hsigma_{s(e)}-\hsigma_{t(e)}+\frac{iQ}{2}\sre)}
{s_b\left(\hsigma_{s(e)}+\hsigma_{t(e)}-\frac{iQ}{2}\sre\right)
s_b\left(\hsigma_{s(e)}-\hsigma_{t(e)}-\frac{iQ}{2}\sre\right)}
\ ,
\label{ZhyperN2}
\eeq
where we defined
\beq
\sre=1-r_e \ .
\label{mudef}
\eeq
In the limit 
\eqref{reduction}, two out of the four factors stays the same
\beq
s_b\left(\pm(\hsigma_{s(e)}-\hsigma_{t(e)})+\frac{iQ}{2}\sre \right)
 \to s_b\left(\pm
(\hsigma_{s(e)}-\hsigma_{t(e)})+\frac{iQ}{2}\sre \right) \ ,
\eeq
whereas the other two factors behave\footnote{The appearance of the
exponential of the quadratic term represents the parity anomaly. This
cancels out in our final expression.}
\beq
s_b\left(\pm(\hsigma_{s(e)}+\hsigma_{t(e)})+\frac{iQ}{2}\sre \right) \to
e^{\pm \frac{i\pi}{2} (\pm (2\sigma+\hsigma_{s(e)}+\hsigma_{t(e)})+\frac{iQ}{2}\sre)^2
} \ .
\eeq
After subtracting the divergences, most of the finite part cancels due
to the $\pm$ sign, and the only remaining term is 
$$
e^{-\frac{\pi Q}{2} (\hsigma_{s(e)}+\hsigma_{t(e)})\sre} .
$$
However, this cancels out when we sum over $e\in E$, because
$$
\sum_{e\in E}(\hsigma_{s(e)}+\hsigma_{t(e)}) \sre=
\sum_{v\in E}\hsigma_v \sum_{e: \textrm{ around } v}  
 \sre = 0 \ .
$$
where we used \eqref{reduction} and that (thanks to \eqref{sumdual2}) 
\beq
\sum_{e: \textrm{ around  }v}\sre=2 \ .
\label{summuzero}
\eeq
The chiralmultiplet 1-loop determinant therefore becomes
\beq
Z_{\rm chiral} \to Z'_{\rm chiral}=
\frac{s_b\left(\sigma_{s(e)}-\sigma_{t(e)}+\frac{i Q}{2} \sre\right)}{s_b\left(\sigma_{s(e)}-\sigma_{t(e)}-\frac{i Q}{2}
\sre\right)} \ .
\eeq

The total partition function still takes the form \eqref{Z3d}, except that
the 1-loop determinants are replaced by $Z'_{\rm vector}, Z'_{\rm
chiral}$.
Using (see \eqref{ebsb})
\beq
\frac{s_b(s+\frac{i Q}{2} r)}{s_b(s- \frac{i Q}{2} r)}
=
e^{Q s \pi r}\frac{e_b(s+\frac{i Q}{2} r)}{e_b(s- \frac{i Q}{2} r)} \  ,
\eeq
we see that this coincides with the Boltzmann weight $W_{\theta}(s)$ 
of the Faddeev-Volkov model \cite{Volkov:1992uv,FaddeevVolkovAbelian,FaddeevCurrent}
in
\cite{Bazhanov:2007mh}
under the parameter identification between 
their rapidity $\theta_e^*$ \footnote{This is denoted by $\theta_e$ in \cite{Bazhanov:2007mh}.}
and the R-charge $\sre$ (recall \eqref{Rchargerule})
\beq
\theta^*_e=\pi \, \sre , \quad
\theta_e:=\pi-\stheta_e=\pi r_e 
\ ,
\label{thetamu}
\eeq
up to an overall normalization\footnote{Our function $e_b$ is denoted
by $\varphi$ in \cite{Bazhanov:2007mh}.}.
Again, the overall normalization worked out in \cite{Bazhanov:2007mh}
is not necessary for our purpose.
This two-step reduction from the solution of the star-triangle
relation in \cite{Bazhanov:2010kz}
to the Faddeev-Volkov model has been explained in a different
manner 
in \cite{Spiridonov:2010em}.

The Faddeev-Volkov model is realizes Virasoro algebra on the lattice,
see also the subsequent formulation of the discrete Liouville theory \cite{Faddeev:2000if}.
As explained in \cite{Terashima:2011qi}, Liouville theory
could be thought of as a boundary theory of $SL(2, \bR)$ Chern-Simons
theory,
and plays crucial roles in the relation with 3d $\scN=2$ gauge theories.
This suggests a more direct relation between the 2d spin system and 
the 3d hyperbolic geometry. 
In the story of \cite{Terashima:2011qi} the geometric picture simplifies
in the semi-classical limit of the Chern-Simons
theory \cite{Terashima:2011xe}.
On the gauge theory side 
this limit is translated into the limit
$b\to 0$, where $S^3_b$ in \eqref{S3b} degenerates to $\bR^2\times S^1_b$
(the radius of $S^1_b$ is small, and is given by $b$).
In the following we are going to take exactly the same limit.
It is worth pointing out that 
the same limit of the Faddeev-Volkov model has been analyzed in the
seminal paper \cite{Bazhanov:2007mh}, which also studied the connection
with circle patterns.

\subsection{Further Reduction to 2d}\label{subsec.2d}

Let us dimensionally reduce our theory further to 2d, by 
taking the $b\to 0$ limit of the $S^3_b$ partition function.
The resulting theory has 2d $\scN=(2,2)$ supersymmetry.
After the dimensional reduction 2d the vector multiplet scalar $\sigma$
is complexified  due to the Wilson lines along $S^1_b$, 
and a 3d real mass parameter reduces to a twisted massed
in 2d.

In this limit, quantum dilogarithms reduce to classical dilogarithms
\eqref{Liclassical},
and we have
\beq
Z'_{\rm 3d}\to Z_{\rm 2d}=\int\! d\sigma\,\, 
\exp\left[-\frac{1}{\pi b^2} \scW_{\rm 2d}(\sigma)\right] \ ,
\eeq
where the effective twisted superpotential $\scW_{\rm 2d}$
is given by
\beq
\scW_{\rm 2d}(\sigma)=
\scW_{\rm 2d}(\rho)=
-\frac{1}{2}
\sum_{e\in E} \left[
  \tilde{l}\left(\rsigma_{s(e)}-\rsigma_{t(e)}+i\stheta_e\right)
-
 \tilde{l}\left(\rsigma_{s(e)}-\rsigma_{t(e)}-i\stheta_e\right)
\right]
\ .
\label{W2d}
\eeq
We have scaled the variable $\sigma$ by $2\pi b$
(this comes the radius of $S^1_b$ as in the standard dimensional reduction)
\beq
\rho_e=2\pi b\, \hsigma_e \ =\pi Q b\,  \sigma_e,
\label{rsigmadef}
\eeq
and $\tilde{l}(x)$ is defined in \eqref{tildeldef} in Appendix. 
Because $\tilde{l}(x+2\pi i)=\tilde{l}(x)$, we can regard $\theta_e$
as defined modulo $2\pi \bZ$. This is consistent with the discussion in 
section \ref{subsec.R-charge}.
Note also the appearance of the classical dilogarithm
represents the sum over the contributions of the KK modes along
$S^1_b$ \cite{Nekrasov:2009uh}\footnote{There 
is a side remark about the integrable structure.
In references \cite{Nekrasov:2009uh,Nekrasov:2009rc},
the saddle point equation of effective twisted superpotential of 
a 2d theory is identified with the
Bethe Ansatz equation of an associated integrable model.
It is interesting to imagine if this integrable structure 
(which in turn is conjectured to be equivalent to the integrable
structure of the dimer model \cite{Goncharov:2011hp})
could be 
related to the integrable structure of the spin system mentioned above.
However, it should be kept in mind that at least the two should not simply be
identified directly; the two spin systems are rather different.
Another difference is that the rapidities in our spin system 
are the chemical potentials for our 4d index,
whereas those in the context of
\cite{Nekrasov:2009uh,Nekrasov:2009rc} are identified with 
the vector multiplet scalars.}.

The classical vacuum of this theory is determined from the saddle point
equation of the twisted superpotential
\beq
\exp\left( \frac{\partial \scW_{\rm 2d}}{\partial \rsigma_v}\right)=1 \ .
\label{W2dN2}
\eeq
In our case, this is given by the ``cross ratio equation''
(which are closely related with the Hirota difference equations)
\beq
1
=\prod_{e\in E,\, s(e)=v}
\frac{
\cosh\left(\rsigma_{s(e)}-\rsigma_{t(e)}+i \stheta_e \right)
}
{
\cosh\left(\rsigma_{s(e)}-\rsigma_{t(e)}- i \stheta_e \right)
}
\prod_{e\in E,\, t(e)=v}
\frac{
\cosh\left(\rsigma_{s(e)}-\rsigma_{t(e)}-i \stheta_e \right)
}
{
\cosh\left(\rsigma_{s(e)}-\rsigma_{t(e)}+ i \stheta_e \right)
}
\ .
\label{Hirota}
\eeq

\subsection{Circle Patterns and Tetrahedra}\label{subsec.tetrahedra}

Surprisingly, the effective twisted superpotential \eqref{W2d}
as well as the saddle point equation \eqref{Hirota} has a beautiful geometrical
reformulation.
For this purpose we need to add extra degrees of freedom to the bipartite graph.
In most of our discussion in previous sections, the bipartite graph only
contains combinatorial data --- we do not need to specify the
length of the edges (the only exception is the part where we discuss
isoradial condition)\footnote{In the context of brane
configurations studied in section \ref{subsec.brane} these
quantitative datum do matter. For example, the area of the region
of the $(N, 0)$-brane represents the inverse square of the gauge
coupling constant. However, the author is not aware of direct brane
interpretation of the geometric construction presented in this section.}.
Henceforth we parametrize the length and angles of the bipartite graph and
zig-zag paths in the following manner.

\begin{figure}[htbp]
\centering{\includegraphics[scale=0.5]{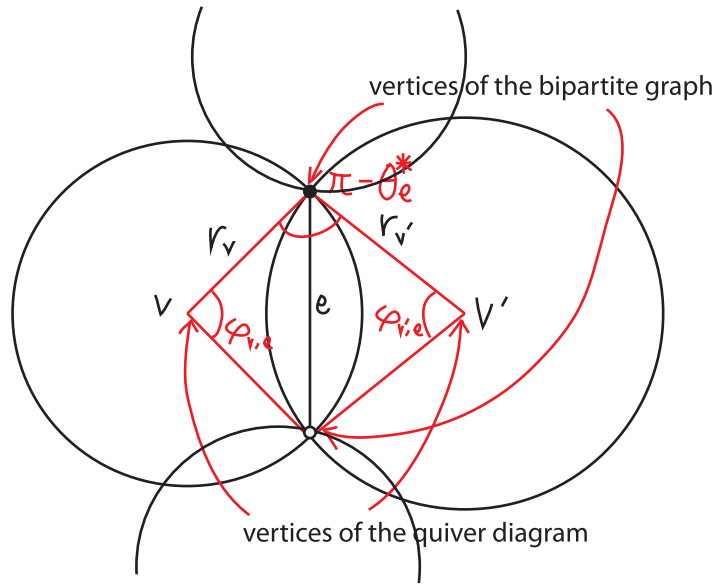}}
\caption{The two circles of radius $r_v$ and $r_{v'}$, each centered at
 vertex $v$ and $v'$ connected with an edge $e$,  intersect at an angle $\stheta_e$. This
 determines the angles $\varphi_{v, e}, \varphi_{v', e}$. The red
 quadrilateral is sometimes called a kite, and is a rhombus for an isoradial circle pattern.}
\label{trigonometrics}
\end{figure}

Let us assign a parameter $\rsigma_v$ at each vertex $v\in V$,
and draw a circle with radius
$r_v:=e^{\rsigma_v}$ centered at $v$.
The quantity $\rsigma_v$ will be identified with the vector multiplet
scalar \eqref{rsigmadef} momentarily, and hence we use the same symbol.
We moreover impose the condition that 
when $v$ and $v'$ are adjacent at an edge $e$
the two circles intersect at an angle
$\stheta_e$.
This angle will again be 
identified $\stheta_e$ introduced previously in \eqref{thetamu}.
Let $\varphi_{v, e}$ to be an angle around vertex $v$ (Figure \ref{trigonometrics}), 
and elementary trigonometry shows that 
\beq
e^{i\varphi_{v, e}}=\frac{e^{\rho_{v}}+e^{\rho_{v'}+i\stheta_e}}{e^{\rho_v}+e^{\rho_{v'}-i\stheta_e}}
=e^{i\stheta_e}\frac{\cosh(\rho_v-\rho_{v'}+
i\stheta_e)}{\cosh(\rho_v-\rho_{v'}-i\stheta_e)}
\label{varphidef}
\ ,
\eeq
see Figure \ref{trigonometrics}.
We also have
\beq
\varphi_{v, e}+\varphi_{v', e}=2\stheta_e \ .
\eeq
Note that the right hand side of this equation takes the same form as
the expression in \eqref{Hirota}.
Thus the cross-ratio equation is the statement that the
angles around $v$ sum up to $2\pi$
\beq
\sum_{e: \, \textrm{around } v} \varphi_{v, e}=2\pi \ ,
\label{varphicond}
\eeq
where we used (see \eqref{summuzero} and \eqref{thetamu})
\beq
\sum_{e: \, \textrm{around } v}\stheta_e\equiv 0 \quad (\textrm{mod}\,
2\pi\bZ) \ .
\label{sumtheta}
\eeq

Summarizing, the saddle point equation \eqref{Hirota} is reformulated as
a geometric condition on a set of circles intersecting at the vertices
of the bipartite graph.
In the literature a set of such circles is called
a circle pattern\footnote{In the case that $\stheta_e=\frac{\pi}{2}$ for all $e\in E$, a
circle pattern reduces to a circle packing together with its dual.}, and the deformation of a circle pattern represents the discrete analog of conformal
transformation.

\begin{figure} 
\centering{\includegraphics[scale=0.5]{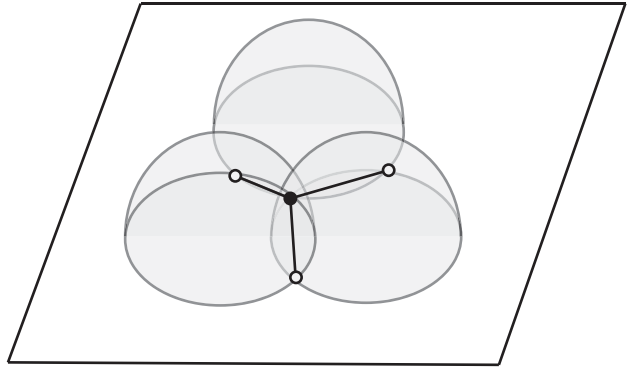}}
\caption{Our polyhedron $M$ is defined as a complement of all the
 hemispheres.}
\label{pattern3D}
\end{figure}

For us, the importance of this geometric reformulation is that it gives
the 3-manifold we are after\footnote{Historically the fact that a circle pattern has to
do with the geometry of 3-manifolds is known since long ago, see for
example \cite[Chapter 13]{ThurstonLecture}.}. 
To explain this,
let us regard $T^2$ as part of the boundary $\bR^2$ of $\bH^3$,
and for each circle consider a hemisphere centered at $v$
and intersecting the boundary $\bR^2$ with
the circle around $v$. 
Let us denote the region above all the hemispheres by
$M$ (Figure \ref{pattern3D}). Note that we still keep the torus identification on the boundary.

This could be equivalently described as follows. For each face $v\in F^*=V$ of the
bipartite graph $\scG^*$ you can associate an ideal polyhedron $\Delta_v$
whose vertices are the vertices of $\scG^*$ at the face and 
at infinity. Then the polygon is defined as the union of all these ideal
polyhedra,
with neighboring faces glued together:
\beq
M=\displaystyle\bigcup_{v\in V} \Delta_v \ .
\eeq
This is a hyperbolic 3-manifold (in the standard metric induced from
that of $\bH^3$), and has a geodesic boundary.
In particular its hyperbolic volume is computed to be
the sum of the volume of $\Delta_v$, and 
depends non-trivially on $\rsigma_v$ and $\theta^*_e$.

\begin{figure}[htbp]
\centering{\includegraphics[scale=0.5]{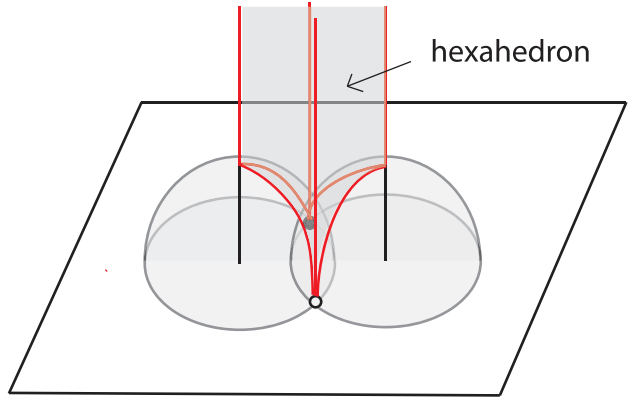}}
\caption{Our polygon $M$ could be decomposed into hexahedra, each of
 which is associated with an edge of the bipartite graph. The projection
 of this hexahedron gives the red quadrilateral
 in Figure \ref{trigonometrics}.}
\label{hexahedron}
\end{figure}

The manifold $M$ can be decomposed
into hexahedra (Figure \ref{hexahedron}), which in turn could be 
divided into two non-ideal tetrahedra.
As is clear from this definition, the gluing condition of the 
hexahedra at the vertices of the quiver diagram
is exactly the condition
\eqref{varphicond}, 
ensuring that the total angle around an edge of the polyhedron is
$2 \pi$. We have therefore shown that the 
vacuum equations of the 2d gauge theory coincides with 
the gluing conditions of the 3-manifold $M$.
As discussed in \cite{Terashima:2012cx}, the existence of the solution of the gluing condition
is guaranteed by the 4d conditions on the R-charge \eqref{sum}, \eqref{sumdual} and the result of \cite{BobenkoS}.

We can also directly compare the value of the twisted superpotential 
with the volume of the hyperbolic 3-manifold. 
We show that
\beq
\scW_{\rm 2d}(\sigma)\big|_\textrm{vacuum}
-\scW_{\rm 2d}\big|_{\sigma=0} =
\Vol[M](\sigma)\big|_\textrm{gluing} - \Vol[M_0] \ ,
\label{2dmain}
\eeq
where the quantity $\Vol[M_0]$, to be defined below, is independent of
$\rho_v$, and we have evaluated the both sides at the values of $\rho_v$
determined from the vacuum equations or gluing conditions.
To evaluate the volume, we use the decomposition of $M$ 
into hexahedra (Figure \ref{hexahedron}).
The volume of the hexahedron at an edge $e$
is given by \cite[Lemma 4.2]{SpringbornThesis}
\begin{align}
V^e_{\rm hexahedron}&
=\Lov\left(\frac{\varphi_{s(e), e}}{2}\right)
+\Lov\left(\frac{\varphi_{t(e), e}}{2}\right) \\
&=
2\Lov\left(\frac{\stheta_e}{2}\right) 
-\frac{1}{2}
\left[
\tilde{l}(\rho_{s(e)}-\rho_{t(e)}+i\stheta_e)-\tilde{l}(i\stheta_e)
\right] \nonumber \\
 &\qquad +\frac{1}{2}
\left[
\tilde{l}(\rho_{s(e)}-\rho_{t(e)}-i\stheta_e)-\tilde{l}(-i\stheta_e)
\right] \label{junk}\\
&
 \qquad +\frac{1}{4}(\varphi_{s(e), e}-\varphi_{t(e), e})(\rho_{s(e)}-\rho_{t(e)}) \nonumber
 \ ,
\end{align}
where in the last line we used \eqref{Lov2}, \eqref{tildelid} in Appendix.
The total volume of $M$ is then given by summing over the edge $e$:
\beq
\textrm{Vol}[M]\big|_{\rm gluing}=\sum_{e\in E} V^e_{\rm hexahedron}\big|_{\rm gluing}
=\scW_{\rm 2d}\big|_{\rm gluing}-\scW_{\rm 2d}\big|_{\sigma=0}+\Vol[M_0] 
\label{volMformula} \ ,
\eeq
where we used that fact that the last term in \eqref{junk} cancels out
at the saddle point (see \eqref{reduction}, \eqref{varphicond}, \eqref{sumtheta}), and we
defined 
\beq
\Vol[M_0]:=\sum_e 2 \Lov\left(\frac{\stheta_e}{2} \right)
\ .
\label{Volisoradial}
\eeq
As we will see shortly this is the value of $\textrm{Vol}[M]$ 
when the circle pattern is isoradial.
This proves \eqref{2dmain}.

\bigskip
\noindent\underline{Isoradiality}

In general, it is a difficult problem to analytically solve the 
cross ratio equations \eqref{Hirota}.
Considerable simplification occurs when the graph is isoradial.
Isoradial condition  (section
\ref{subsec.quivers}) states that it is geometrically 
possible to choose $\rho_v$ such that 
\beq
\rho_v=\rho \quad \textrm{independent of }v \ .
\eeq
This is particularly useful for us because 
in this case \eqref{varphidef} simplifies to 
\beq
\varphi_e\equiv \theta^*_e  \quad (\textrm{mod}\,\, 2\pi \bZ) \ , 
\label{phitheta}
\eeq 
which automatically solves
\eqref{varphicond} (see \eqref{sumtheta}).  
There is an enhanced $SU(|V|)$ global symmetry at this vacuum.

In this vacuum the volume of $M$ reduces to \eqref{Volisoradial},
and the critical value of the twisted superpotential
vanishes,
\beq
\scW_{\rm 2d}(\sigma)\Big|_{\rm isoradial} =0 \ .
\eeq
Note that the isoradiality is not preserved under the 
two fundamental moves of Figure \ref{moves}. However, the critical value of the
twisted superpotential is still preserved under the fundamental moves.

\bigskip
\noindent\underline{Hyperbolic Circle Patterns}

To this point 
we started with a geometric realization of the dimensionally reduced theory after the Higgsing. This raises a natural question:
what if we start with the 3d theory before Higgsing and dimensionally
reduce to 2d? Does this theory has geometric reformulation? The answer
to the latter question is affirmative. Let us explain this briefly.

Let us go back to the expression of the partition function 
in section \ref{subsec.3d}, and again let us take $N=2$,
in which case the chiralmultiplet 1-loop determinant is 
given in \eqref{ZhyperN2}. After the dimensional reduction as in section \ref{subsec.2d} (but without going through the Higgsing in section 
\ref{subsec.Abelianization}), the 2d vacuum equation is written as
\beq
\sum_{e: \textrm{ around  }v} \tilde{\varphi}_{v,e}=2\pi,  
\label{newgluing}
\eeq
where $\tilde{\varphi}_{v,e}$ is defined by
\beq
e^{2i
\tilde{\varphi}_{v, e}}=
\frac{1+e^{\rho_{v',e}-\rho_{v,e}+i\theta_e}}{1+e^{\rho_{v',e}-\rho_{v,e}-i\theta_e}}
\frac{1+e^{\rho_{v,'e}+\rho_{v,e}-i\theta_e}}{1+e^{\rho_{v',e}+\rho_{v,e}+i\theta_e}} \ ,
\label{newgluing2}
\eeq
where the edge $e$ connects the vertex $v$ with another $v'$.
The equation \eqref{newgluing2}
is a trigonometric relation for the angle placed on Figure \ref{trigonometrics},
but now placed on a {\it hyperbolic} plane $\bH^2$ instead of $\bR^2$ \cite{SpringbornThesis}. From this viewpoint the Higgsing in section \ref{subsec.Abelianization} is a process of replacing circle patterns in $\bH^2$ by those in $\bR^2$.

\subsection{Comments on the Geometry of the 3-manifold}\label{subsec.M5}

In the previous subsection our 3-manifold is obtained by gluing ideal
polyhedra. Here are 
preliminary remarks on the geometry 
of this 3-manifold; more details are currently under investigation.

\bigskip
\noindent\underline{M5-brane Realizations}

Let us comment on the brane realization of our 3d $\scN=2$ theory.
We begin with the brane configuration in section \ref{subsec.brane},
and in order to dimensionally reduce the theory to 3d we compactify 
and T-dualize
along the $3$-direction.
Then we have a D4/NS5 system in type IIA:
\beq
\begin{tabular}{c||c|c|c|c|c|c|c|c|c|c}
 &  0 & 1& 2 & 3& 4& 5& 6& 7& 8& 9  \\
\hline
$N$ D4 &  -- & -- & -- &   & &  $\cdot$ &  &  $\cdot$ & &  \\
\hline
$1$ NS5 &  --& --& -- & -- & \multicolumn{4}{c|}{$\Sigma$} & &  
\end{tabular}
\eeq
This 
can be lifted
to M-theory by including an extra $11$-th direction:
\beq
\begin{tabular}{c||c|c|c|c|c|c|c|c|c|c|c}
 &  0 & 1& 2 & 3& 4& 5& 6& 7& 8& 9  & 11\\
\hline
$N$ M5 &  -- & -- & -- &   & &  $\cdot$ &  &  $\cdot$ & & & -- \\
\hline
M5 &  --& --& -- & -- & \multicolumn{4}{c|}{$\Sigma$} & & &  
\end{tabular}
\eeq
The resulting system is the intersection of two different types of M5-branes,
where $N$ M5-branes (coming from D5-branes in type IIB) wraps $T^3$
and another M5-brane (coming from the NS5-brane) wraps a special
Lagrangian
submanifold $M'$ in $\bR^3\times T^3$. Note that $M'$ and $T^3$, 
two different types of M5-branes, 
intersect along a one dimensional subspace\footnote{Similar brane configuration has been studied in \cite{Lee:2006hw}.}.
This brane configuration tells us that $M'$ is the Riemann surface fibered
over the $S^1$; of course, this $S^1$ is part of the $S^1\times S^3$
on which we compactified 4d gauge theory.
This places a strong constraint
on the geometry of $M'$. It is a future problem to clarify the precise relation between $M'$ and our 3-manifold $M$.

\bigskip
\noindent\underline{$M$ As a Link Complement}\label{subsec.knot}

Our 3-manifold $M$ is a link complement,
at least in some special situations. Let us illustrate this with a simple example, following \cite[Appendix]{LackenbyAgolThurston} and \cite{PurcellIntroduction}.

\begin{figure}[htbp]
\centering{\includegraphics[scale=0.5]{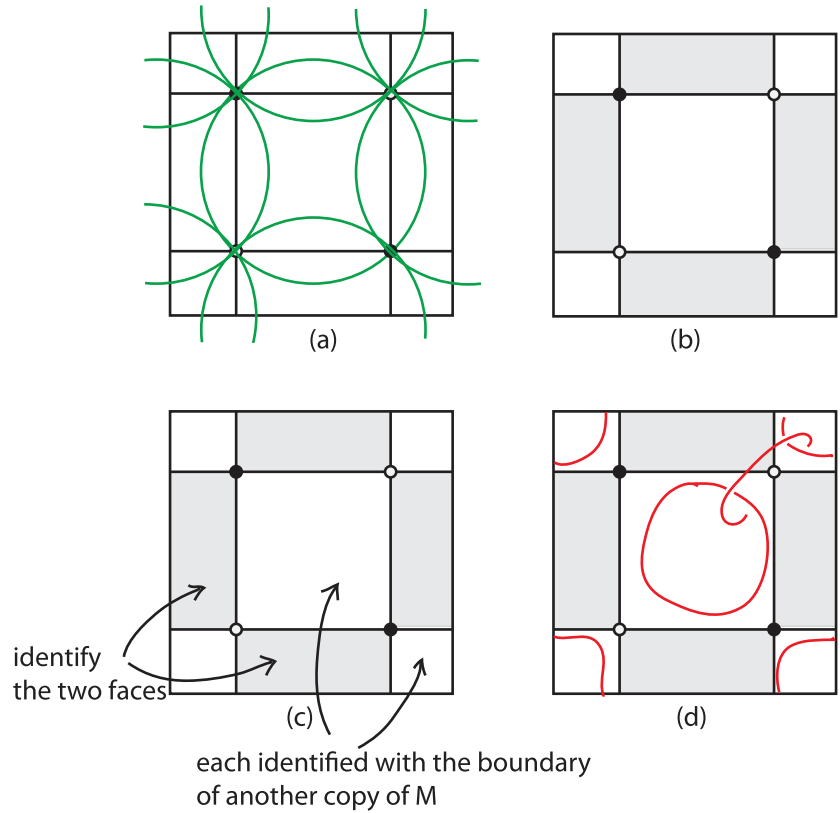}}
\caption{From two identical copies of our 3-manifold $M$ we obtain the
 link complement shown in (d). We first start with a circle packing and
 its dual (a). We then choose a checkerboard coloring of the faces (b),
and then pairwise glue the colored faces. This leaves a polyhedron with
 uncolored faces as boundaries, together with its copy.
We finally glue the two polyhedra along the uncolored faces, 
and obtain a link complement in (d).
}
\label{linkeg}
\end{figure}

Suppose that we start with a bipartite graph shown in Figure
\ref{linkeg} (a).
The corresponding Calabi-Yau geometry is the canonical bundle over $\bP^1\times \bP^1$.
Suppose that the R-charge of all the fields are $1/2$. 
The circle pattern then is a circle packing with its dual (Figure
\ref{linkeg} (a), green circles).
The bipartite graph $\scG^*$ have four faces (gauge groups), and we choose a checkerboard
coloring of the faces\footnote{This coloring is different from the
coloring described in section \ref{subsec.zigzag}.} (Figure
\ref{linkeg} (b)).
We first glue two colored faces. The resulting manifold has a
boundary.
We next prepare an identical copy of this partially glued polyhedron
and then glue the two along the remaining uncolored faces. 
This gives a link complement, shown in Figure \ref{linkeg} (d).

More generally, this procedure works if 
(1) a circle pattern is given by a circle packing with its dual, {\it
i.e.,}
when the R-charges are all canonical,
(2) the vertices of the quiver diagram can be checkerboard colored
and (3) choose a particular pairing of the black faces.
For example, this works for all the orbifolds of the example in Figure \ref{linkeg} with canonical R-charges.

\section{Dimers and BPS State Counting}\label{sec.dimers}

In this section we study the relations of our results to BPS state
counting and topological string theory.

\subsection{Thermodynamic Limit of Dimers}\label{subsec.thermodynamics}

Let us first begin with the observation that the
Legendre transform of the volume of the
3-manifold 
$M$ when the circle pattern in isoradial coincides
with the thermodynamic limit of a dimer partition function.

Suppose that I have a dimer model on $T^2$.
We choose a parametrization of the weight $e^{\nu_e}$
for an edge $e\in E$ by an angle (``rhombus angle'')
$\stheta_e$:
\beq
e^{\nu_e}=2\sin\frac{\stheta_e}{2} \ .
\label{thetaedef}
\eeq
Note that this is the logarithm of the length of the edge $e$ in Figure \ref{trigonometrics}.
The partition function of this dimer model is defined by
\beq
Z_1=\sum_{m: \textrm{ perfect matching}} \prod_{e \in m} \, e^{\nu_e} \ ,
\eeq
where the sum is over perfect matchings, {\it i.e.,} subsets of edges
of the bipartite graph containing each vertex once and exactly once.

Let us consider the thermodynamic limit of this partition function. 
As is standard in statistical mechanics, this limit is taken by 
enlarging the size of the torus by $n$ times 
in the two directions of the torus.
Let $Z_n$ be the partition function of this model. In the large $n$
limit, the free energy scales as $n^2$, and we could extract a 
finite quantity by taking the scaling limit. In the case that the dimer is
isoradial,
this is computed
to be \cite{KenyonLaplacian,Tiliere}\footnote{Our $\stheta_e$ here differs from $\theta$ in \cite{KenyonLaplacian, Tiliere} by a factor of $2$.}
\beq
2 \pi \lim_{n\to \infty}
\frac{1}{n^2} \log Z_n
= \sum_{e\in E}
\left(
    \stheta_e \log 2 \sin \frac{\stheta_e}{2} +  
2\Lov\left(\frac{\stheta_e}{2}\right) \right) =:\tilde{V}(\nu) \ .
\label{thermlim1}
\eeq
This quantity is originally 
introduced as a normalized determinant of the discrete
Dirac operator \cite{KenyonLaplacian}. 
This is consistent with the fact that dimers are described by free
fermions,
and the dimer partition function is the determinant 
of the Dirac operator.

The expression \eqref{thermlim1} is not exactly the volume itself
\eqref{Volisoradial},
but its Legendre transform.
In fact, there is a general analysis of the thermodynamic limit
of the dimer model in \cite{KenyonOkounkovSheffield},
which is applicable to arbitrary (not necessarily isoradial) dimers. 
There it has been shown that the Legendre transformation of the 
thermodynamic limit of the dimer partition function (the left hand side
of \eqref{thermlim1})
is the Ronkin function $R$ (explained in Appendix \ref{app.ronkin})
of the spectral curve of the dimer model,
and is given by the Legendre transformation of the surface tension $\sigma$:
\beq
\tilde{V}(\nu)=\sum_{e\in E} \stheta_e \nu_e + \sigma (\stheta) \ ,
\label{KOSLegendre}
\eeq
where $\stheta_e$ is the conjugate variable to $\nu_e$.
By definition $\stheta_e$ can be computed as the probability of the edge
$e$
to be chosen in a perfect matching, and is shown to be related to
$\stheta_e$
by \eqref{thetaedef} \cite{KenyonLaplacian}.

This fact is powerful enough to determine $\sigma(\stheta)$.
In fact, from the property of the Legendre transformation we have
\beq
\frac{\partial \sigma(\stheta)}{\partial \stheta_e}=-\nu_e=-\log
2\sin\frac{\stheta_e}{2} \ ,
\eeq
and by integrating with respect to $\stheta_e$ (see \eqref{Li2})
we obtain
\beq
\sigma(\stheta)=\sum_{e\in E} 2\Lov\left(\frac{\stheta_e}{2}\right) =\Vol[M_0] \ .
\label{sigmaexpr}
\eeq
By combining \eqref{KOSLegendre} and \eqref{sigmaexpr}
we obtain \eqref{thermlim1}.
In Appendix \ref{app.ronkin} we demonstrate
\eqref{thermlim1} by an explicit computation for a simple example.

Summarizing, we have seen that the Legendre transform of 
the volume of the polygon $M$ coincides with the thermodynamic limit of
the dimer partition function.

\bigskip
\noindent
\underline{Comments on F-extremization}

Before proceeding to a next subsection
there is a side remark on IR anomalous dimensions.
The combination \eqref{thermlim1}
can be rewritten as
\begin{equation}
\begin{split}
&\stheta \log \left(2\sin \frac{\stheta}{2}\right)+2\Lov\left(\frac{\stheta}{2}\right) \\
& \qquad \qquad =
-2 \pi \left[-
z \log(1-e^{2\pi i z})+\frac{i}{2}\left(
\pi z^2+\frac{1}{\pi}\textrm{Li}_2(e^{2\pi i z})\right)
-\frac{i\pi}{12} 
\right]\ ,
\end{split}
\end{equation}
where $\stheta= 2\pi  z$ and the function inside the bracket is the function
$l(z)$ defined in \cite{Jafferis:2010un}, which appears in the 
1-loop determinant for the $S^3_{b=1}$ partition function.
This means that the extremization 
of the dimer partition function coincides with the 
extremization of the integrand of the $S^3_{b=1}$
partition function 
in \cite{Jafferis:2010un},
and hence the IR anomalous dimension
in the saddle point approximation of the matrix model\footnote{
A caveat in this statement is that there is a 3d global 
symmetry which is anomalous and not included in the 
superconformal index in 4d.}.
The author is not aware of an immediate application of this fact,
nevertheless it would be interesting to explore
if there is any further implication of this
observation.

\subsection{Relation with Topological Strings}\label{subsec.topstring}

We have seen that our partition function 
coincides with the thermodynamic limit of the dimer partition function. 
The same dimer partition
function
appears in the physics of BPS state counting and 
topological string theory (see \cite{Yamazaki:2010fz,Sulkowski:2011qs} for summary).

Let us consider type IIA string theory on the 
toric Calabi-Yau $X_{\Delta}$. We consider BPS bound states of 
D0/D2-branes wrapping 0/2-cycles of $X_{\Delta}$
bound to a single D6-brane filling the whole $X_{\Delta}$\footnote{
We can define the
crystal melting model with D4-branes wrapping 4-cycles included,
however
its thermodynamic limit has some additional subtlety due to the
existence of the ``gas phase'' \cite{KenyonOkounkovSheffield}.}. 
Equivalently these bound states are 1/2 BPS particles of
the 4d $\scN=2$ supersymmetric gauge theory.

We define the BPS partition function as the generating function of the
BPS degeneracies of these bound state of D-branes
(mathematically this is the generating function for the 
generalized Donaldson-Thomas invariants):
\beq
Z_{\rm BPS}(g_s, t)=\sum_{p_0, p_A} \Omega(p_0, p_A) e^{-g_{\rm top} p_0 -t_A
p_A} \ ,
\eeq
where $p_0$ and $p_A$ are D0 and D2-brane charges, respectively,
and $\Omega$ is the BPS degeneracy.

In the case the Calabi-Yau 3-fold is toric, then the BPS partition
function
is known to be described by the
partition function of the crystal melting model, or equivalently a dimer
model on $\bR^2$
\cite{Szendroi:2007nu,Mozgovoy:2008fd,Ooguri:2008yb}.
The resulting partition function takes an infinite product form,
and is a reduction of a square of the topological string
partition function \cite{Aganagic:2009kf}\footnote{Interestingly, 
this partition function could also be expressed as a matrix model \cite{Ooguri:2010yk,Szabo:2010sd},
which is again written in terms of dilogarithm functions and is
similar to the matrix models considered in subsections 
\ref{subsec.3d} and \ref{subsec.Abelianization}.
The spectral curve of this matrix model reproduces the mirror curve.}.

The thermodynamic limit of the dimer partition function in the previous
subsection
appears in the thermodynamic limit of this dimer model, {\it i.e.,}
in the limit $g_{\rm top}\to 0$. This limit has been analyzed in
\cite{KenyonOkounkovSheffield},
and the limit (after suitable regularization)
is an integral of the Ronkin function $R(x,y)$
of the spectral curve of the
dimer model,
which as shown in \cite{Ooguri:2009ri}
coincides with the Riemann surface appearing in the 
mirror of the Calabi-Yau manifold \eqref{mirrorCY3}.
This leads to the identification
\beq
Z_{\rm BPS}\to \exp\left[\frac{1}{g_s^2} \scF_0 \right]  , \quad
\scF_{{\rm top, }0}=\int \! dx dy \, R(x,y)  \ ,
\eeq
where $\scF_{{\rm top, }0}$ is the genus $0$ prepotential 
of the topological B-model on the mirror $\check{X}_{\Delta}$,
or equivalently of the topological A-model on $X_{\Delta}$,
and the integral is over the amoeba defined in Appendix \ref{app.ronkin}.
Combining these with our result \eqref{Volisoradial}
we see that an integral of the Legendre transform of the critical value of the
effective twisted superpotential reproduces the genus $0$
topological string partition function, as stated in \eqref{VolLegendre}.
\beq
\scF_{\textrm{top, }0}= \int \! dx dy \, \, \scL \left[\textrm{Vol}[M] \Big|_\textrm{isoradial} \right]\ ,
\eeq
where $\scL$ represents the Legendre transformation described in section \ref{subsec.thermodynamics}.

As a simple check of this relation, let us count the number of parameters.
As we have seen, the twisted superpotential $\scW_{\rm 2d}(\sigma)$,
 and therefore $\Vol[M](\sigma)$, 
have $d-1$ parameters.  Because we integrate over two of them, we have 
$d-3$ remaining parameters. This is the same as the number of compact 
$P^1$ in the geometry, and hence the number of K\"ahler moduli, for a 
toric Calabi-Yau 3-folds without compact 4-cycles.

\bigskip
\noindent\underline{4d/1d correspondence}

We have seen that a reduction of the 4d superconformal index for quiver gauge
theories reproduces the 
topological string partition function on the dual geometry, 
the toric Calabi-Yau manifold.
As we have explained above the two quantities are apparently of very
different origin: one comes from 4d $\scN=1$ superconformal quiver {\it gauge
theory}
and another from the counting of BPS {\it particles} in toric Calabi-Yau.
This raises the obvious question whether there is a natural explanation
for this correspondence.

The simple answer is that the same bipartite graph (quiver and the superpotential)
describes the two different physics, 
(1) 4d quiver gauge theory and (2) 1d quantum mechanics on the
1/2
BPS particles inside 4d $\scN=2$ theory\footnote{This was 
one of the key ingredients which lead to the construction of crystal melting
model \cite{Ooguri:2008yb}.
The 
subtle difference is that 
in (2) we have a D6-brane filling the whole $X_{\Delta}$.
This plays crucial roles in the wall crossing phenomena,
but not for the consideration of this section.}.
In this description, Seiberg duality in 4d mutates
the quiver, which on the 1d side is interpreted as the crossing of the
wall of marginal stability in the moduli space
\cite{Chuang:2008aw,Aganagic:2010qr}. 
Because the 4d index is invariant under Seiberg duality,
we should recover a quantity which is not affected by wall crossing, 
and topological string partition function precisely satisfies this criterion.

This correspondence between 4d and 1d 
is strongly reminiscent of the correspondence
\cite{Cecotti:2011iy}
between the 3d $\scN=2$ gauge theories realized as domain walls in 4d $\scN=2$ theories and 
the 1/2 BPS particles in the same 4d $\scN=2$ theory,
where the counterpart of a 4d Seiberg duality is played by a 3d mirror symmetry.

\section{Concluding Remarks}\label{sec.conclusion}

We conclude with a few open problems, 
besides those already mentioned in introduction.

Our results, in particular the relation \eqref{I4d=Z2d}, are reminiscent of the recently found relation between 4d
$\scN=2$
superconformal index and 2d TQFT \cite{Gadde:2009kb,Gadde:2011ik,Gadde:2011uv}. 
There are many important differences
between the two proposals; we have different class of 4d
SCFTs (for example, our theories have Lagrangians whereas their theories
do not in general) with different amount of supersymmetry ($\scN=1$ versus $\scN=2$).
However, it is instructive to pursue the analogy between the two.
For example, the counterpart of the decomposition of 4d $\scN=2$ SCFT 
from the pants decomposition of the Riemann surface \cite{Gaiotto:2009we}
is the decomposition of the 4d $\scN=1$ SCFT from the partial resolution
of the toric Calabi-Yau singularities, and the mirror of the resolution
is indeed the pants
decomposition of the mirror curve \eqref{mirrorcurve}.

The relation between the 2d spin system and the 3d $SL(2)$ Chern-Simons
theory deserves further study. Since 3d $SL(2)$ Chern-Simons theory is
closely related with 3d gravity, our 2d spin system could be regarded as a kind of 
spin network deconstructing 3d gravity.

In the discussion of 3-manifolds in this paper, 
the natural starting point was the 4d superconformal index.
At the technical level this means that the quantum dilogarithm function
arises as a limit of the elliptic gamma function. 
Since quantum dilogarithm function appears extensively 
in the geometry of 3-manifolds, wall crossing phenomena
and 3d $\scN=2$ gauge theories\footnote{
The underlying mathematical structure is the cluster algebras,
see \cite{Goncharov:2011hp} for relation with dimer models.
The same mathematical structure appears recently 
in the study of scattering 
amplitudes \cite{ArkaniHamedTalks}, which also utilize zig-zag paths
on bipartite graphs.
}, it is natural to ask
if we could lift all these to the level of elliptic gamma functions.
For example, is there a pentagon relation for elliptic gamma functions?

A somewhat related problem is to give a mathematical formulation of the
superconformal index. Since our 4d theory is defined purely from the
toric data there should be a mathematical formulation of the
superconformal index in term of the geometry of the toric Calabi-Yau
3-fold, perhaps as an equivariant character of some moduli space\footnote{
This is related to the computation of the index on the gravity dual. However,
our superconformal index does not capture the IR R-charge and  hence
should be formulated in the language of the toric Calabi-Yau cone, rather than of the Sasaki-Einstein manifold.}.
Pushing this further, there should be ``categorified'' versions of 
\eqref{I4d=Z2d}, \eqref{Z3dZ3d}, for example 
\beq
\scH^{\rm BPS}_\textrm{4d $\scN=1$ SCFT on $S^3$}=\scH_\textrm{moduli space
defined from $X_{\Delta}$} \ , 
\eeq
for some Hilbert space $\scH$ (similar categorified viewpoint
has been advocated in \cite{Terashima:2011qi}).
Our results in section \ref{sec.dimers} suggest a relation of this
Hilbert space with the moduli space of ideal sheaves on the Calabi-Yau manifold.

\section*{Acknowledgments}
The author would like thank Yuji Terashima for collaboration on a companion paper \cite{Terashima:2012cx}
and for pointing out important references
which are crucial for the completion of this work.
He also thanks Tudor D. Dimofte, Leo P. Kadanoff, Sangmin Lee, Jessica
S. Purcell, Salvatore Torquato, Grigory
S. Vartanov, Dan Xie
and Herman L. Verlinde for stimulating discussion
and correspondence.
He would like to thank the Bogoliubov Laboratory of Theoretical Physics 
(JINR, Dubna), Newton Institute (Cambridge University) 
and Ludwig-Maximilians-Universit\"{a}t M\"{u}nchen for
hospitality where part of this work has been performed.
The results of this paper are presented at the conference 
``New perspectives on supersymmetric gauge theories''
and the high energy theory seminar at Princeton University, and the
author 
thanks the 
audience for feedback.
Finally, he thanks Princeton Center for Theoretical Science 
for its generous support.

\appendix

\section{Special Functions}\label{sec.special}

In this appendix we summarize formulas for special function used in
this paper. 

We define the elliptic gamma function $\Gamma(z;p,q)$
by\footnote{Beaware of the notational differences in comparison with the
literature. For example, our notation here is different
from that in \cite{Bazhanov:2010kz}.}
\beq
\Gamma(z;p,q)=\prod_{j,k=0}^{\infty}
\frac{1-z^{-1} q^{j+1}p^{k+1}}{1-z q^{j} p^{k}} \ .
\label{ellGammadef}
\eeq
This is the basic building block for the 4d superconformal index.
This function satisfies
\beq
\Gamma(z;p,q)=\frac{1}{\Gamma(pq/z;p,q)} \ ,
\label{Gammainverse}
\eeq
and
\beq
\Gamma(pz;p,q)=\theta(z;q)\Gamma(z;p,q), \quad
\Gamma(qz;p,q)=\theta(z;p)\Gamma(z;p,q) \ ,
\label{Gammadiff}
\eeq
where $\theta(x;p)$ is the $q$-theta function defined by
\beq
\theta(z;q)=(z;q)_{\infty} (q/z;q)_{\infty} \ ,
\label{qtheta}
\eeq
with
\beq
(a;q)_{\infty}=\prod_{k=0}^{\infty}(1-a q^k) \ .
\label{infbracket}
\eeq
$q$-theta function is related to the Jacobi $\theta$-function 
\beq
\theta_1(z|q)
=2 q^{1/4}\sin \pi z \prod_{m=1}^{\infty}(1-q^{2m})(1-2\cos 2\pi z
q^{2m}+q^{4m}) \ ,
\label{Jacobitheta}
\eeq
by
\beq
\theta_1(z|q)= i q^{1/4}(q^2;q^2)_{\infty} e^{-\pi i z} \, \theta(e^{2\pi i z};q^2) \ .
\label{thetarelation}
\eeq
The equation \eqref{Gammadiff} is a elliptic generalization of the
familiar relation $\Gamma(x+1)=x \Gamma(x)$.
From \eqref{Gammainverse}, \eqref{Gammadiff}, \eqref{qtheta} it follows that
\beq
\frac{1}{(1-z)(1-z^{-1})\Gamma(z;p,q)\Gamma(z^{-1};p,q)}=
\frac{1}{(1-z)^2} \theta(z;p)\theta(z;q) \ .
\label{usefulgamma}
\eeq

As for the 3d and 2d partition functions, we need classical and quantum dilogarithms.
The classical dilogarithm function (Euler dilogarithm) $\Li_2(z)$ is 
defined by
\beq
\textrm{Li}_2(z)=-\int_0^z \frac{\log(1-t)}{t}dt =\sum_{n=1}^{\infty}
\frac{z^n}{n^2} \ . 
\label{Li2}
\eeq
A related function, Lobachevsky function $\Lov(x)$, is defined by
\beq
\Lov(x)=-\int_{0}^x \! dt \, \log \big| 2\sin t \big|\ .
\eeq
This could be expressed in terms of the Euler dilogarithm to
be
\beq
2i \Lov(\theta)=\textrm{Li}_2(e^{2 i\theta})-\textrm{Li}_2(1)-\theta(\theta-\pi)
\ ,
\label{LovLi}
\eeq
where the famous formula by Euler states that
\beq
\textrm{Li}_2(1)=\zeta(2)=\frac{\pi^2}{6} \ .
\label{zeta2}
\eeq
By definition the Lobachevsky function is an odd function,
$\Lov(-x)=-\Lov(x)$.
Correspondingly, Euler dilogarithm satisfies
\beq
\Li_2(-e^x)+\Li_2(-e^{-x})=-\frac{\pi^2}{6}-\frac{1}{2}x^2 ,
\label{Li2sum}
\eeq
which is consistent with \eqref{zeta2}.
We can also show
\beq
\Lov\left(\frac{\pi-\theta}{2}\right)=\Lov\left(\frac{\theta}{2}\right)-\frac{1}{2}\Lov\left(\theta\right)
\label{Lov2}
\eeq

We also need the non-compact quantum
dilogarithm functions $s_b(z)$ and $e_b(z)$ \cite{FaddeevVolkovAbelian,FaddeevKashaevQuantum,Faddeev95}.
These functions are defined by
\beq
s_b(z)=\exp\left[ \frac{1}{i} \int_0^{\infty} \frac{dw}{w} \left(
\frac{\sin 2zw}{2\sinh (bw) \sinh (w/b)}-\frac{z}{w} \right) \right] \ ,
\label{sbdef}
\eeq
and
\beq
e_b(z)=\exp\left(\frac{1}{4} \int_{-\infty+i0}^{\infty+i0}  \frac{dw}{w}
\frac{e^{-i 2zw} }{\sinh(wb) \sinh(w/b)}\right) \ , 
\label{ebdef}
\eeq
where the integration contour in \eqref{ebdef} is chosen above the pole
at $w=0$. 
In both these expressions we require $|\mathrm{Im}\, z| <|\mathrm{Im}\, c_b|$ for convergence at infinity. There is a simple relation between the two functions 
\beq
e_b(z)=e^{\frac{\pi i z^2}{2}} e^{- \frac{i\pi (2-Q^2)}{24}}s_b(z) \ ,
\label{ebsb}
\eeq
and we loosely refer to both functions as quantum dilogarithms.

The function $s_b(z)$ satisfies a difference equation
\beq
s_b\left(z-\frac{ib^{\pm 1}}{2} \right)=2 \cosh \left(\pi b^{\pm 1} z\right)
s_b\left(z+\frac{ib^{\pm 1}}{2}\right) \ ,
\label{sbdiffeqn}
\eeq
and hence
\beq
s_b\left(z-\frac{iQ}{2} \right)=4 \sinh \left(\pi b^{- 1} z\right)
 \sinh \left(\pi b z\right)
s_b\left(z+\frac{iQ}{2}\right) \ .
\label{sbdiffeqn2}
\eeq
We can use these equations to analytically continue 
$s_b(z), e_b(z)$ to the whole complex plane. 
The position of poles and the zeros are represented as
\beq
s_b(z)=\prod_{m,n\in \bZ_{\ge 0}}
\frac{mb+nb^{-1}+\frac{Q}{2}-iz}{mb+nb^{-1}+\frac{Q}{2}+iz} \ .
\label{sbpole}
\eeq
where the left hand side is to be interpreted as a regularization of the
right hand side.
By definition we have
\beq
s_b(z)=s_{1/b}(z) \ ,
\label{sbselfdual}
\eeq
and
\beq
s_b(z) s_b(-z)=1 \ .
\label{sbinverse}
\eeq
The asymptotic limit is given by
\beq
s_b(z)\to e^{\textrm{sgn}(z)\frac{i\pi z^2}{2}} , \quad \textrm{as }
z\to \pm \infty \ .
\label{sblimit} 
\eeq

The name ``quantum dilogarithm'' could be justified by the fact that in the classical limit
$b\to 0$, 
the quantum dilogarithm reduces to the classical dilogarithm:
\beq
e_b(z)\to \exp \left[\frac{1}{2 \pi b^2} 
  (-i)\textrm{Li}_{2}(-e^{2\pi b z}) \right], \quad
s_b(z)\to \exp\left[\frac{1}{2\pi b^2} \tilde{l}(2\pi b z)\right] \ ,
\label{Liclassical}
\eeq
where
\begin{equation}
\begin{split}
\tilde{l}(z):&=(-i)\left[\textrm{Li}_2(-e^z)+\frac{1}{4}z^2+\frac{\pi^2}{12}\right] 
=\frac{-i}{2} \left[\textrm{Li}_2(-e^z)-\textrm{Li}_2(-e^{-z}) \right]
 \\ 
& = 2\Lov\left(\frac{z}{2i}+\frac{\pi}{2}\right)
\ .
\end{split}
\label{tildeldef}
\end{equation}
This function is odd, and its derivative is given by
\beq
\tilde{l}'(z)=i\log 2\cosh \frac{z}{2} \ .
\eeq
In the main text we need the following equality
\beq
\frac{1}{2}
\left[
  \tilde{l}(\rho+i\theta)-\tilde{l}(\rho-i\theta)
\right]
= \Lov(\theta)-\Lov\left(\frac{\varphi_1}{2}\right)
-\Lov\left(\frac{\varphi_2}{2}\right)
+\frac{1}{4}\rho(\varphi_2-\varphi_1)
\ ,
\label{tildelid}
\eeq
where $\varphi_1, \varphi_2$ are defined by
\beq
e^{i\varphi_1}= \frac{1+e^{\rho+i\theta}}{1+e^{\rho-i\theta}},
\quad
e^{i\varphi_2}= \frac{1+e^{-\rho+i\theta}}{1+e^{-\rho-i\theta}},
\quad 
\varphi_1+\varphi_2=2\theta 
\ .
\eeq
The equality \eqref{tildelid} can be proven with the help of the
formula (this is a minor modification of \cite[Proposition A]{Kirillov:1994en})
\beq
\Li_2(-e^{\rho +i\theta})-\Li_2(-e^{\rho -i\theta})
= 2 i \left[\Lov(\theta)-
\Lov(\omega)-\Lov(\theta-\omega)-\omega\rho
\right] \ ,
\eeq
with
\beq
\tan\omega=\frac{e^{\rho}\sin \theta}{1+e^{\rho}\cos\theta} \quad\textrm{i.e.,}\quad e^{2i
\omega}=\frac{1+e^{\rho+i\theta}}{1+e^{\rho-i\theta}}\ .
\eeq

\section{Hyperbolic Volume and Ronkin Function}\label{app.ronkin}

In this appendix we first recall why the thermodynamic limit of the
dimer partition function is given by the Ronkin function of the 
characteristic polynomial of the dimer model. 

We then show in an
explicit example that the Ronkin function is the Legendre transform of
the hyperbolic volume of the polyhedron $M$. 
This equality is known in the literature, for example for $\bC^3$ \cite{CerfKenyon}
and for a certain parametrization for the canonical bundle over
$\bP^1\times\bP^1$ \cite{CohnProppKenyon}.

Consider combinations of $\nu_e$ along closed non-trivial loops
($\alpha$- and $\beta$-cycles) in $T^2$, and 
let us denote them by 
$e^x, e^y$. These are the chemical potentials for the so-called ``height
function'' of the dimer model.
Written in this variable, the partition function for the enlarged dimer
$Z_n$ (defined in section \ref{subsec.thermodynamics})
is simply given by
\beq
Z_n(e^x, e^y)=\prod_{z'^n=e^x}\prod_{w'^n=e^y} Z_1(z', w') \ ,
\eeq
where only the dependence of $Z_n$ with respect to $x, y$ are explicitly
shown here. By taking a logarithm and dividing by $n^2$, the product
becomes a Riemann sum and we have
\begin{align}
\lim_{n\to \infty} \frac{1}{n^2} \log Z_n(e^x, e^y)
=
\oint_{|z'|=|w'|=1} \frac{dz'}{z'} \frac{dw'}{w'} \log \big| Z_1(e^x z', e^y w')
\big| =R_{Z_1}(x, y) \ ,
\label{thermlim2}
\end{align}
where for a Laurent polynomial $P(z, w)$
we defined its Ronkin function $R_{P}$
to be
\beq
R_{P}(x,y)=\frac{1}{(2\pi i)^2} \oint_{|z|=|w|=1} \frac{dz}{z} 
\frac{dw}{w} \log
\big| P(e^x z,e^y w) \big| \ .
\eeq
This is a convex function, and is linear outside the amoeba $\scA_P$,
defined by
\beq
\scA_P=\{(x,y)\in \bR^2 \big| \exists (\theta, \phi)\in (\bR/2\pi
\bZ)^2, \, 
P(e^{x+i\theta}, e^{y+i\phi})=0 \} \ .
\eeq
In general it is not easy to work out the exact analytic expression for
the Ronkin function. However, in many cases it is much simpler to
compute its derivative by taking the residues of the integral.

Let us check the equality of \eqref{thermlim1} and 
\eqref{thermlim2} (times $\pi$)
when $\Delta$ is the toric diagram for $\bC^3$.
The dimer has three edges, whose weights we denote by 
$x, y, z$. 
Correspondingly we have three rhombus angles $\alpha, \beta, \gamma$,
which is related to $x, y, z$ by
$$
e^x=2\sin \alpha, \quad e^y=2 \sin \beta, \quad e^z=2 \sin \gamma \ .
$$
and satisfying
$$
\alpha+\beta+\gamma = \pi \ .
$$
These weights satisfy the triangular inequality 
(we take $\alpha, \beta, \gamma \ge 0$)
\beq
e^x+e^y \ge e^z, \quad e^y+e^z\ge e^x, \quad e^z+e^x \ge e^y 
\ .
\label{triangular}
\eeq
The dimer partition function is 
\beq
Z_1=e^x+e^y+e^z \ ,
\eeq
whose amoeba coincides with the region determined by \eqref{triangular}.
We now show 
\beq
\pi
R_{Z_1}(x, y, z)=\alpha  x+\beta y+\gamma
z+\Lov(\alpha)+\Lov(\beta)+\Lov(\gamma) \ .
\label{C3check}
\eeq
inside amoeba,
where we defined the 3-variable version of Ronkin function by
\beq
R_{P}(x,y,z)=\frac{1}{(2\pi i)^3} \oint_{|u|=|v|=|w|=1} \frac{du}{u} 
\frac{dv}{v} \frac{dw}{w}\log
\big| P(e^x u,e^y v, e^z w) \big| \ .
\eeq
This coincides with the 2-variable Ronkin function $R_{1+\frac{y}{x}+\frac{z}{x}}$
up to a constant factor.

To check \eqref{C3check} let us compute the derivative of the both sides of \eqref{C3check}.
The derivative of the left hand side
inside amoeba 
can be evaluated by taking
residues \cite{Fujimori:2008ee}
\beq
\pi \frac{\partial R_{Z_1}}{\partial x}(x, y, z)=
\pi \frac{\partial R_{1+e^x+e^y}}{\partial x}(x-y, z-y)
=\pi-\cos^{-1}\left(
\frac{e^{2(x-y)}-e^{2(z-y)}-1}{2 e^{z-y}}
\right)  \ ,
\eeq
and hence by the law of cosines
\beq
\pi\frac{\partial R_{Z_1}}{\partial x}=\pi-\cos^{-1}\left(
-\cos\alpha \right)
=\alpha \ .
\eeq
The computation is similar for the $y, z$-derivatives,
and the result coincides with the $x , y, z$-derivatives 
of the right hand side of 
\eqref{C3check}.
Note that the amoeba coincides with the region determined by \eqref{triangular}.


\bibliographystyle{JHEP}
\bibliography{index3mfdbib}

\end{document}